\documentclass[12pt,a4paper]{elsarticle}

\usepackage{graphicx}
\usepackage{amsmath,amssymb}
\usepackage{slashed}

\biboptions{sort&compress}

\newcommand{\vev}[1]{\langle #1 \rangle}
\newcommand{\td}{t_d}
\newcommand{\tkin}{t_{kin}}
\newcommand{\tthr}{t_{thr}}
\newcommand{\athr}{a_{thr}}
\newcommand{\Gt}{{\widetilde{G}}}
\newcommand{\gt}{{\tilde{g}}}

\newcommand{\GeV}{\,\textrm{GeV}\,}
\newcommand{\OR}{\,\textrm{or},\qquad}
\newcommand{\calA}{\mathcal{A}}
\newcommand{\half}{\frac{1}{2}}
\newcommand{\paren}[1]{\left( #1 \right)}

\newcommand{\Sect}[1]{Section #1} \newcommand{\sect}[1]{section #1}

\newcommand{\Eq}{Eq.\ } \newcommand{\eq}{eq.\ }
\newcommand{\eqs}{eqs.\ }

 \newcommand{\figs}{figures }

\newcommand{\Tab}{Table }

\newcommand{\refe}{ref.\ } \newcommand{\Refe}{Ref.\ }
\newcommand{\refes}{refs.\ }

\begin{document}

\title{Kinetic and chemical equilibrium of the Universe and gravitino
production}
\author[prl]{Raghavan Rangarajan}
\ead{raghavan@prl.res.in}
\author[prl]{Anjishnu Sarkar\fnref{1}}
\ead{anjishnu@lnmiit.ac.in}
\address[prl]{Physical Research Laboratory, Navrangpura, Ahmedabad 380009,
India}
\fntext[1]{Current address: The LNM Institute 
of Information Technology, Jaipur 302031, India.}

\date{\today}

\begin{abstract}
Flat directions in generic supersymmetric theories can change the thermal
history of the Universe. A novel scenario was proposed earlier where the
vacuum expectation value of the flat directions induces large masses
for all the gauge bosons and gauginos. This delays the thermalization
of the Universe after inflation and solves the gravitino problem. In
this article we perform a detailed calculation of the above scenario.
We include the appropriate initial state particle distribution functions,
consider the conditions for the feasibility of the non-thermal scenario,
and investigate phase space suppression of gravitino production in
the context of heavy gauge bosons and gauginos in the final state. 
We find that the total gravitino abundance generated is consistent 
with cosmological constraints.
\end{abstract}


% \arxivnumber{}

\maketitle
 
\section{Introduction}
\label{sec:intro} 

Large vacuum expectation values (vevs) for flat directions in
supersymmetric (SUSY) theories, i.e., directions in the field space
of scalars that have a flat potential, can affect the thermal history
of the Universe in two ways as discussed in the literature. Firstly,
they may delay the thermalization of the inflaton decay products
\cite{Allahverdi:2005fq,Allahverdi:2005mz,Allahverdi:2010xz}.  Secondly,
the flat directions could potentially dominate over the inflaton decay
products \cite{Olive:2006uw,Gumrukcuoglu:2008fk}, such that reheating is
associated with the decay of flat directions as opposed to the decay of
the inflaton. Below we shall consider gravitino production in the context
of flat directions delaying thermalization of the decay products of the
inflaton which are presumed to dominate the Universe.

Gauge symmetries are broken along flat directions and for certain
flat directions all standard model (SM) gauge symmetries are
broken in the early Universe. (Low energy phenomenology is not
affected as the vevs of these flat directions ultimately disappear.)
In \refes \cite{Allahverdi:2005fq,Allahverdi:2005mz} the authors have
presented an interesting scenario in which flat directions give a
large mass $\sim \sqrt{\alpha} \varphi$ to all gauge
bosons and gauginos where $\varphi$ is the vev of the field $\phi$
that parametrizes the flat direction and $\alpha = g^2/(4\pi)$ for gauge
coupling constant $g$. Since processes involving gauge
bosons are important for thermalization, this results in a slow rate of
thermalization after the end of inflation. Thus the Universe enters a
period of pre-thermalization after inflation. Throughout this article
we use the term thermalization to refer to chemical equilibration,
and thermal equilibrium to refer to chemical equilibrium.

In the scenario of \refes \cite{Allahverdi:2005fq,Allahverdi:2005mz}
the Universe reaches a state of kinetic equilibrium via $2 \rightarrow 2$
interactions after a certain time interval has elapsed following inflaton
decay.  Further thermal equilibrium is reached only when number violating
interactions, i.e., $2 \rightarrow 3$ interactions, are also effective.
Thus the rate of inflaton decay, of kinetic equilibration and of thermal
equilibration follow the relation
\begin{equation}
\Gamma_d \gg \Gamma_{kin} > \Gamma_{thr} \,,
\end{equation}
where
\begin{align}
\Gamma_d &= \textrm{Inflaton decay rate} \nonumber \\
\Gamma_{kin} &= \textrm{Rate of kinetic equilibration} \nonumber \\
\Gamma_{thr} &= \textrm{Rate of thermal equilibration,} 
\label{eq:raterel} 
\end{align}
and we define $t_d$, $\tkin$ and $\tthr$ as the times of inflaton decay, 
kinetic equilibration and chemical equilibration respectively.

In \refes \cite{Allahverdi:2005fq,Allahverdi:2005mz} the authors argue
that due to the delay in thermalization the generic problem of gravitino
overproduction in SUSY models is also avoided. The lack of thermalization
leads to a dilute plasma of high energy particles, and the low number
density of particles effectively leads to low gravitino production.
Moreover due to a low final reheat temperature after thermalization
gravitino production after thermalization is also suppressed.

In this article we provide a detailed analysis of gravitino
production in the non-thermal Universe scenario of \refes
\cite{Allahverdi:2005fq,Allahverdi:2005mz}.  To find the number density
of gravitinos $n_\Gt$, \refes \cite{Allahverdi:2005fq,Allahverdi:2005mz}
use the integrated Boltzmann equation of the form
\begin{equation}
\dot{n}_\Gt + 3 H n_\Gt = \sum_{I \leq J} \vev{\sigma v_{rel}} ~n_I n_J ,
\label{eq:stdbltzeq} 
\end{equation}
from inflaton decay till thermal equilibration for the process $I +
J \rightarrow \Gt + K$, where $I$ and $J$ are quarks or squarks and $K$
is a gauge boson or a gaugino.  We too obtain and use this equation, but
only for $\tkin<t<\tthr$ where one can describe the particle distribution
function by a thermal distribution with an effective chemical potential.
For $t<\tkin$ we instead use a delta function distribution function for
the incoming particles and then evaluate the collision integral for the
integrated Boltzman equation.  As part of our analysis, we have also
explicitly obtained the conditions for obtaining a non-thermal Universe
after inflaton decay.

There is an important effect associated with large vevs of the flat
directions as relevant for gravitino production that has not been
considered earlier.  Gauge bosons and gauginos are produced along
with gravitinos in the relevant processes for gravitino production. As
mentioned above these get large masses $m_{\gt,g}$ due to the large vevs
of the flat directions.  Consequently gravitino production ceases in
cases where the energy of quarks and squarks in the Universe falls below
the mass of gauge bosons and gauginos due to phase space suppression.

As in ref. \cite{Allahverdi:2005mz} we consider two cases, $m_0<\Gamma_d$ 
and $m_0>\Gamma_d$, where $m_0$ is the condensate mass.  We find that for
$m_0<\Gamma_d$ phase space suppression shuts off gravitino production both
before thermalization and even after thermalization, till the condensate
decays.  For $m_0>\Gamma_d$, gravitino production is
shut off after thermalization due to 
phase space suppression till the condensate decays.

It should be noted here that the longevity of these flat directions has
been in dispute \cite{Allahverdi:1999je,Postma:2003gc,Olive:2006uw,
Allahverdi:2006xh,Basboll:2007vt,Basboll:2008gc,Allahverdi:2008pf,
Gumrukcuoglu:2008fk,Gumrukcuoglu:2009fj,Allahverdi:2010xz}.  One of the
key issues is whether or not non-perturbative effects lead to a fast decay
of the condensate.  In \refe \cite{Allahverdi:1999je,Postma:2003gc} it
was pointed out that non-perturbative decay due to parametric resonance
is suppressed for a complex scalar condensate if the real and imaginary
parts of the scalar field oscillate out of phase.  In that case the field
does not pass through the minimum of the potential at $\varphi=0$ and so
the required non-adiabatic condition $|\dot{\omega}_k|\sim \omega_k^2$ is
not satisfied, where $\omega_k$ is the the frequency of the $k$-th mode of
the $\chi$ field produced by the decay of the condensate.  (Backreaction
effects can also be important in suppressing non-perturbative decay.)
The condensate then decays only perturbatively and hence lasts for a
long time. \Refe \cite{Olive:2006uw} however argued that in the case of
multiple flat directions there is a mixing of excitations along different
directions which makes non-perturbative effects important.  Further
investigations in \refe \cite{Allahverdi:2006xh} implied that even with
multiple flat directions resonant decay is most likely to be suppressed
(in cases of physical interest) because the multiple flat directions
effectively reduce to one flat direction or a collection of independent
single flat directions. Raising certain issues related to gauge fixing,
\refes \cite{Basboll:2007vt,Basboll:2008gc} worked in the unitary gauge
to eliminate the Nambu-Goldstone boson that appears in the context of SUSY
flat directions charged under the gauge group of the MSSM and studied the
circumstances under which the presence of multiple flat directions can
lead to non-perturbative decay.  \Refe \cite{Allahverdi:2008pf} invoked
charge conservation to argue that non-perturbative decay is suppressed.
At the same time \refe \cite{Gumrukcuoglu:2008fk} analysed the issue
with proper gauge fixing and concluded that with one flat direction
there is no resonant particle production, in agreement with \refe
\cite{Allahverdi:2008pf}, while disagreeing with the conclusion of \refe
\cite{Allahverdi:2008pf} in the context of multiple flat directions.
However \refes \cite{Allahverdi:2008pf,Allahverdi:2010xz} argue that
even if non-perturbative decay occurs for multiple flat directions it
leads to redistribution of energy of the condensate amongst the fields
in the D flat superspace and hence to practically the same cosmological
consequences (including large gauge boson and gaugino masses and delayed
thermalization) as in the scenario with only perturbative decay.

Our analysis below presumes that perturbative decay is the relevant decay
mechanism of the flat direction condensate and that the impact of the 
condensate lasts long enough to have cosmological consequences.  This may
be because either there is only a single flat direction that is
excited, or because the multiple flat directions are excited such that
parametric resonance is still suppressed, or because of the arguments
of \refes \cite{Allahverdi:2008pf,Allahverdi:2010xz}, or because of
backreaction effects.

There are other effects that affect the evolution and decay of
the flat direction condensate. Thermal contributions to the
condensate potential can change the mass of the condensate
leading to an earlier oscillation time for the condensate
\cite{Allahverdi:2000zd,Anisimov:2000wx,Anisimov:2001dp}. Below we take
the condensate mass to be of order of the SUSY breaking scale $m_0 = 100
\GeV$. We present our results in terms of $m_0$ which can be replaced by
larger values to include these effects. (\Refe \cite{Anisimov:2001dp}
obtains masses of $\sim 10^{10}\GeV, 10^{6} \GeV$ and $10^5 \GeV$ for
$n = 1,2,3$ where terms that lift the flat directions have the form
$\phi^{2n+4}/M_*^{2n}$ where $M_* = 10^{18} \GeV$.) Another effect is
that the condensate can decay due to scattering by inflaton decay products
\cite{Dine:1995kz,Allahverdi:2000zd, Anisimov:2000wx}. Both these effects
have been discussed in Appendix A.4 of \refe \cite{Allahverdi:2005mz} where
it has been argued that these effects will not be relevant during the
non-thermal phase.

The flat direction condensate can also decay to Q-balls due to
inhomogeneities in the condensate field. This has been studied in the
context of both gauge mediated and gravity mediated SUSY breaking
\cite{Kusenko:1997si,Enqvist:1997si,Kasuya:1999wu,Kasuya:2000wx,
Enqvist:2000gq,Kasuya:2000sc,Enqvist:2000cq,Kasuya:2001hg,Multamaki:2002hv}.
In addition it has been seen that condensate lumps with different
properties than Q-balls called Q-axitons may also be formed
\cite{Enqvist:1999mv}. Note however that the time scale for the
formation of Q-balls and Q-axitons can be large. For gauge mediated
SUSY breaking scenarios the time scale for Q-ball formation $\tau
\sim 5 \times 10^{5} \,m^{-1}$ \cite{Kasuya:1999wu}, where $m$ is a
characteristic mass scale for the flat direction condensate, and for
gravity mediated SUSY breaking scenarios $\tau \sim 100-1000 \,m^{-1}$
\cite{Enqvist:1997si} or $5.5 \times 10^{3} m^{-1}$ \cite{Kasuya:2000wx}.
\Refe \cite{Kusenko:1997si} also obtains a Q-ball formation time which
is larger than the flat direction oscillation time by a factor of $1.5
\times 10^4$.  The time scale for Q-axiton formation is $1600 \,m^{-1}$
\cite{Enqvist:1999mv}. Q-ball formation may be avoided for certain flat
directions, including those with a large stop admixture, which allow for
$K > 0$, where $K$ is a parameter in the potential in gauge mediated
SUSY breaking scenarios \cite{Enqvist:2000gq}. Our analysis does not
include Q-ball or Q-axiton formation.

We now present a summary of this article.  In \sect{\ref{sec:boltzmann}}
we provide the integrated Boltzmann equation and phase space distribution
functions valid for a non-thermal Universe.  \Sect{\ref{sec:revproc}}
provides the scattering processes relevant for gravitino
production in the scenario with heavy gauge bosons and gauginos.
In \sect{\ref{sec:discuss}} we obtain the relevant time scales, namely,
$\tkin$, $\tthr$, $t_f$ and $t_G$ where $t_f$ is the time of decay of
the flat direction condensate and $t_G$ is the time in the radiation
dominated era after thermalization when gravitino production could
commence. In this section we also discuss the feasibility of gravitino
production in the context of phase space suppression in different epochs.
The cross section for gravitino production in the center-of-mass
frame is obtained in \sect{\ref{sec:sigmaCM}}.  Then in
\sect{\ref{sec:Gtabund}} we obtain the gravitino abundance generated in
different time intervals in the Universe.  The parameters required to
obtain the gravitino abundance are given in \sect{\ref{sec:params}}. Our
results are given in \sect{\ref{sec:results}} and our conclusions in
\sect{\ref{sec:conclusions}}.  The Appendices contain derivations of
certain expressions used in the text.

\section{The Boltzmann equation}
\label{sec:boltzmann} 

To obtain the number density of a species $X_3$ participating in reactions
$X_1 X_2 \rightleftharpoons X_3 X_4$ the integrated Boltzmann equation
is
\begin{align}
\dot{n}_3 + 3 H n_3 &=
- \int ~d\pi_1 ~d\pi_2 ~d\pi_3 ~d\pi_4 ~(2\pi)^4
\delta^4(p_1 + p_2 - p_3-p_4)
\nonumber \\ & 
\qquad \times [ f_3 f_4 - f_1 f_2 ]
|\mathcal{M}|^2 \,,
\label{eq:bltz2t2} 
\end{align}
where $f_i$ are phase space distribution functions and
\begin{equation}
d\pi_i \equiv \frac{g_i}{(2\pi)^3} \frac{d^3p_i}{2E_i} \,. 
\end{equation}
$g_i$ is the number of internal degrees of freedom of species $i$.
The $|\mathcal{M}|^2$ includes an average over initial and final internal 
degrees of freedom \cite{Kolb:1988aj}.

Typically $X_1$ and $X_2$ are assumed to be in full thermal equilibrium
with zero chemical potential and thus the phase space distribution
function has the form
\begin{equation}
f_i = \exp(-E_i/T), \qquad i = 1,2. 
\end{equation}
This is because $X_1$ and $X_2$ usually have other stronger
interactions than their interaction with $X_3, X_4$. This
assumption is not valid in the scenario considered in \refes
\cite{Allahverdi:2005fq,Allahverdi:2005mz} as the Universe is not
in thermal equilibrium before $t_{thr}$. In fact $f_{1,2}$ have the
following form
\begin{subequations}
\label{eq:fis}
\begin{alignat}{3}
f_i &= C_i \delta\left(E_i - \frac{m_\phi}{2} \frac{a_d}{a} \right)
        &\qquad& \textrm{ for } \td < t < \tkin 
\label{eq:fi1} 
\\[3mm]
    &= \exp \left(-\frac{E-\xi_i}{T} \right)
        &\qquad& \textrm{ for } \tkin < t < \tthr
\label{eq:fi2} 
\end{alignat}
\end{subequations}
where,
\begin{align}
C_i &= \frac{\rho_\phi (\td)}{m_\phi^3} \left( \frac{a_d}{a}\right)   
    \frac{(2\pi)^3}{\pi g_i},
\qquad i = 1,2
\end{align}
as derived in \ref{sec:phspace}. In the above expressions
$m_\phi$ is the mass of the inflaton, $\rho_\phi(t_d)$ is the  energy
density of inflaton $\phi$ at time $\td$, $a$ is the scale factor
and $\xi_i$ is the effective chemical potential.  \Eq \eqref{eq:fi1}
reflects the scenario that before $\tkin$ the relativistic energy of
fermions and sfermions (whose interactions produce gravitinos) is their
energy at the time of their production from (2 body) inflaton decay at
$\td$ scaled by $1/a$. Between $\tkin$ and $\tthr$ one does not have full
thermalization as represented by $\xi_i$ in \eq \eqref{eq:fi2}. (Also
see footnote 24 of \refe \cite{Allahverdi:2005mz}.)

When the number density of $X_3$ is small, as in our case where $X_3$
represents the gravitino, we can ignore the product $f_3 f_4$ in
\eq \eqref{eq:bltz2t2}. Thus, \eq \eqref{eq:bltz2t2} can be written as
\begin{align}
\dot{n}_3 + 3 H n_3 &= \int ~d\Pi_1 ~d\Pi_2 
~f_1 ~f_2 ~W_{12}(s) 
\equiv A,
\label{eq:mbltzeq} 
\end{align}
where we have defined a variable $A$ to denote the collision integral on
the right-hand-side of the above equation. Here, $W_{12}(s)$ is 
dimensionless and Lorentz invariant, and dependent only on the Mandelstam
variable $s$. $W_{12}$ is given by
\begin{align}
W_{12}(s) &= 
    \int ~d\Pi_3 ~d\Pi_4 ~(2\pi)^4 
        \,\delta^4(p_1+p_2-p_3-p_4) ~|\mathcal{M}|^2 \,,
\label{eq:w12}
\end{align}
which can also be written as $4 E_1 E_2 \,\sigma v_{rel}$ where $v_{rel}$ is
$[(p_1.p_2)^2 - m_1^2 m_2^2]^{1/2}/(E_1 E_2)$.

\subsection{The collision integral}
\label{ssec:colint}

For the completeness of the paper we provide below some details on how
to obtain the collision integral.  It is easier to calculate $W_{12}$
of \eq \eqref{eq:mbltzeq} in the centre-of-mass (CM) frame and it can
then be expressed as \cite{Edsjo:1997bg}
\begin{equation}
W_{12}(s) =
4 p_{12} ~\sqrt{s} ~\sigma_{CM}(s) \,,
\label{eq:w12cm} 
\end{equation}
where
\begin{equation}
p_{12} = \frac{\left[ s - (m_1+m_2)^2 \right]^{1/2} 
\left[ s- (m_1-m_2)^2 \right]^{1/2}}{2\sqrt{s}} 
\label{eq:p12} 
\end{equation}
is the magnitude of the 
momentum of particle $X_1$ (or $X_2$) in the center-of-mass frame
of the particle pair $(X_1,X_2)$.  The cross section in the CM frame
$\sigma_{CM}$ is
\begin{equation}
\sigma_{CM}(s) = \int d\Omega \left( \frac{d\sigma}{d\Omega}
\right)_{CM} \,,
\label{eq:scatt} 
\end{equation}
where
\begin{align}
\left.\frac{d\sigma}{d\Omega}\right|_{CM} &= 
\frac{g_3 g_4}{64 \pi^2 s} \left[ \frac{(s-(m_3+m_4)^2) (s-(m_3-m_4)^2)}
{(s-(m_1+m_2)^2) (s-(m_1-m_2)^2)} \right]^{1/2}
|\mathcal{M}|^2
\label{eq:diffscatt} 
\end{align}
is the differential scattering cross section  in the CM frame
\cite{Lahiri:2005sm}.  
% \cite{Lahiri:2005smch7}.  
% \cite{[{}][{(see Ch. 7)}]Lahiri:2005sm}.  
Note that the averaging over the final state
degrees of freedom in $|\mathcal{M}|^2$ cancels the prefactor $g_3 g_4$ in
the above equation.  $|\mathcal{M}|^2(s,t,u)$ in \eq \eqref{eq:diffscatt}
can expressed in terms of $s$ and $\theta_{13}$, the angle between $X_1$
and $X_3$ in the CM frame, using $u = -s -t + \sum_i m_i^2$ and $t =
m_1^2 + m_3^2 - 2 E_1 E_3 + 2 \sqrt{E_1^2 -m_1^2} ~\sqrt{E_3^2 -m_3^2}
~\cos\theta_{13}$, with $E_1 = \sqrt{p_{12}^2 + m_1^2}$ and $E_3 =
(s + m_3^2 - m_4^2) / (2 \sqrt{s})$ 
\cite{Lahiri:2005sm}.
% \cite{Lahiri:2005smch7}.
% \cite{[{}][{(see Ch. 7)}]Lahiri:2005sm}.  

To perform the integration over $d\Pi_1 ~d\Pi_2$ in (\ref{eq:mbltzeq}) we
note that
\begin{equation}
\int ~d\Pi_1 ~d\Pi_2 \equiv \int ~\left(\frac{g_1}{(2\pi)^3}
~\frac{d^3 p_1}{2E_1}\right)
~\left(\frac{g_2}{(2\pi)^3} ~\frac{d^3 p_2}{2E_2}\right),
\end{equation}
and that the volume element $d^3 p_1 ~d^3 p_2$ is given by 
\cite{Edsjo:1997bg},
\begin{equation}
d^3 p_1 d^3 p_2 = 4\pi |\vec{p}_1| E_1 ~dE_1 ~4\pi |\vec{p}_2| E_2 ~dE_2
 \,\frac{1}{2} \,d(\cos\theta_{21}) \,, 
\end{equation}
where $\theta_{21}$ is the angle between $X_1$ and $X_2$. (Note that
we now work in the rest frame of the Universe.) In order to simplify
the integration we replace the variable $\cos\theta_{21}$ by $s$ via
the relation $s = m_1^2 + m_2^2 + 2 E_1 E_2 - 2|\vec{p}_1||\vec{p}_2|
\cos\theta_{21}$.  The lower limit of the integral over $s$ should not
be less than either $(m_1+m_2)^2$ or $(m_3+m_4)^2$ as dictated by the
phase space factor in \eq \eqref{eq:diffscatt}.  Therefore
\begin{equation}
\int d\Pi_1 ~d\Pi_2 = \frac{g_1 \,g_2}{4 (2\pi)^4} 
\int_{m_1}^\infty \,dE_1 \int_{m_2}^\infty \,dE_2
\int_{s_{min}}^{s_2} \,ds ,
\label{eq:dP1dP2} 
\end{equation}
where $s_{min} = \textrm{max} ~[s_1, (m_1+m_2)^2, (m_3+m_4)^2]$
and $s_{1,2}=m_1^2 + m_2^2 + 2 E_1 E_2 \mp 2|\vec{p}_1||\vec{p}_2|$.
For $m_1, m_2, m_3 \approx 0$, $s_{min} \approx m_4^2$.  Using \eq
\eqref{eq:dP1dP2} in \eq \eqref{eq:mbltzeq} we get
\begin{equation}
\dot{n}_3 + 3 H n_3 = \frac{g_1 \,g_2}{4 (2\pi)^4} 
\int_{m_1}^\infty \,dE_1 \int_{m_2}^\infty \,dE_2 
\int_{s_{min}}^{s_2} \,ds \,f_1 \,f_2 \,W_{12} .
\label{eq:dn3dt} 
\end{equation}

Thus, after calculating $|\mathcal{M}|^2$ for each process mentioned
in the next section, we follow the procedure given above to find
the parameter $W_{12}(s)$ and then with the $f_i$ as given in
\eq \eqref{eq:fis} we use the above equation to calculate the number
density of gravitinos produced between $t_d$ and $t_{kin}$, and $t_{kin}$
and $t_{thr}$.

\section{Relevant processes for gravitino production}
\label{sec:revproc}

\begin{figure}[!tp]
\begin{center}
\begin{tabular}{ccc}
\includegraphics[width=0.4\textwidth]{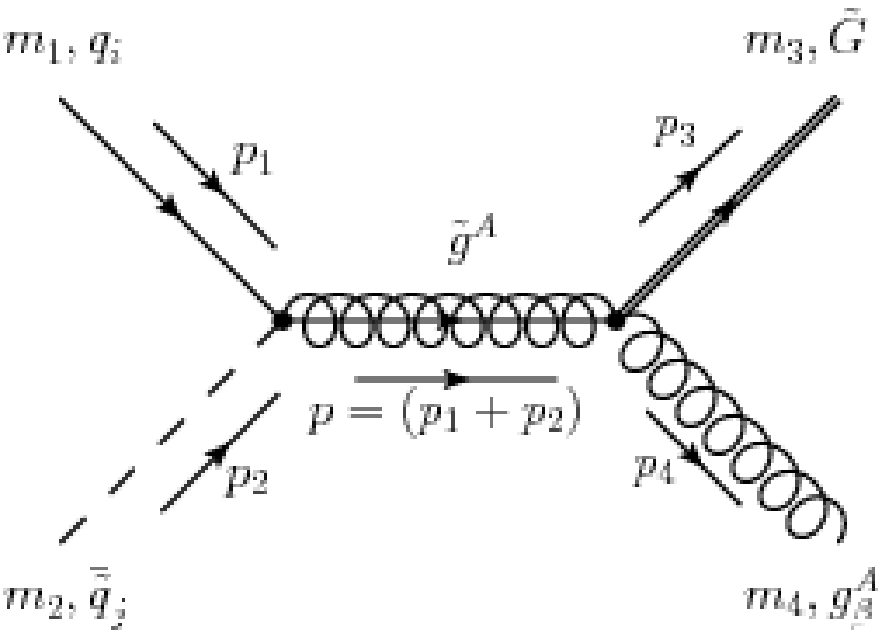}
&\qquad&
\includegraphics[width=0.4\textwidth]{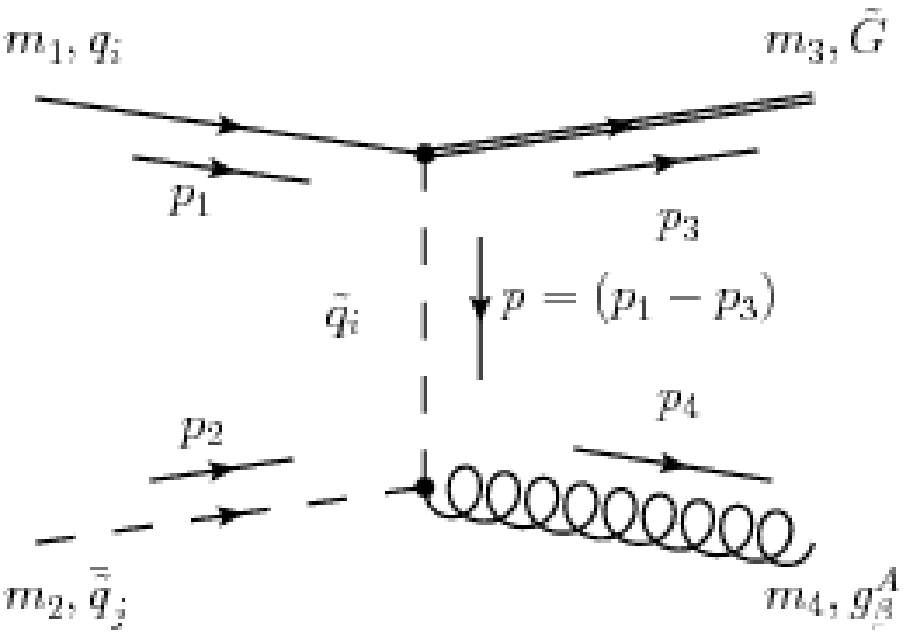}\\ 
(a) && (b) \\ [0.5cm]
\includegraphics[width=0.4\textwidth]{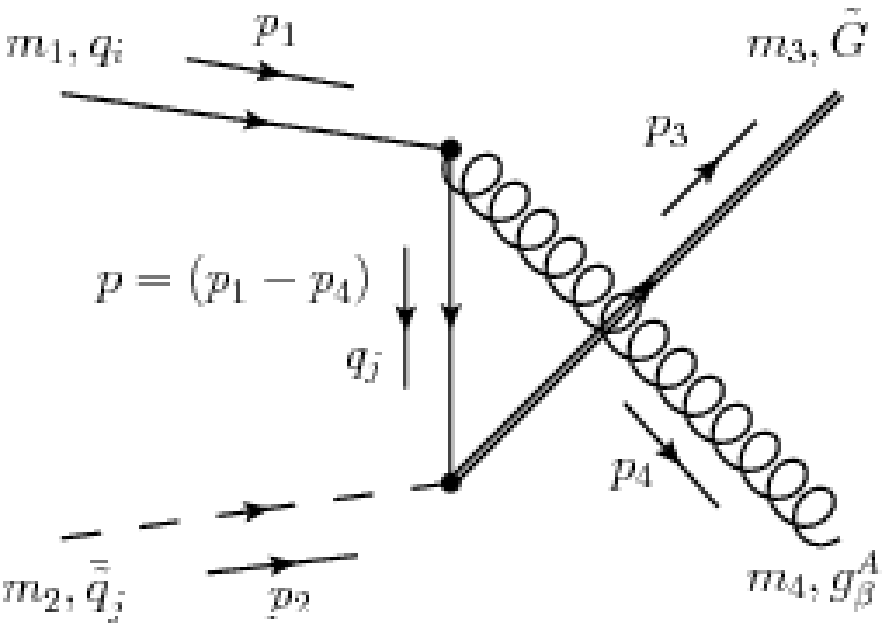}
&\qquad&
\includegraphics[width=0.4\textwidth]{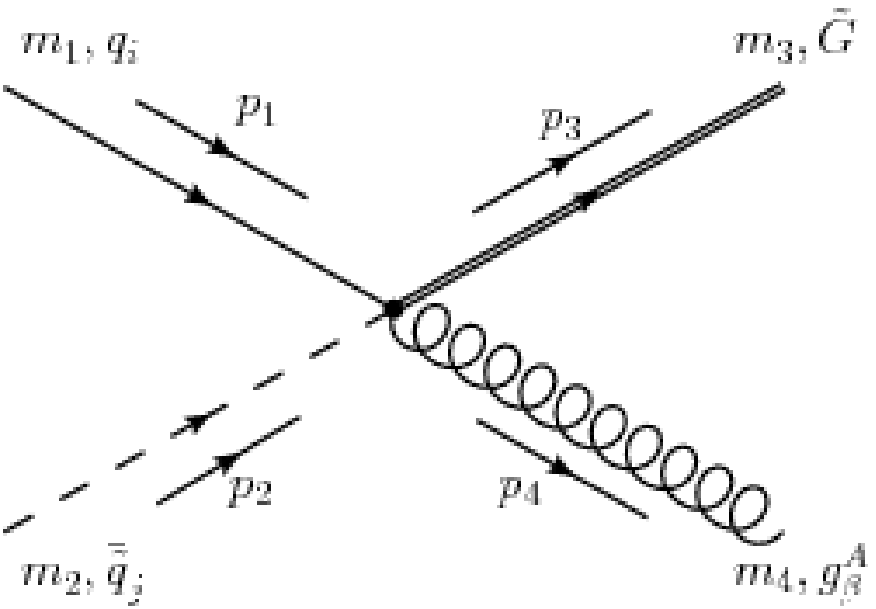}\\ 
(c) && (d) 
\end{tabular} 
\end{center}
\caption{Tree level diagrams for the E process: 
		$q_i + \bar{\tilde{q}}_j 
    \rightarrow g^A + \tilde{G}$ 
    scattering.  The diagrams for
    $\bar{q}_i + \tilde{q}_j  \rightarrow g^A + \tilde{G}$ are similar.
}
\label{fig:e-diags}
\end{figure}

\begin{figure}[!htp]
\begin{center}
\begin{tabular}{ccc}
\includegraphics[width=0.3\textwidth]{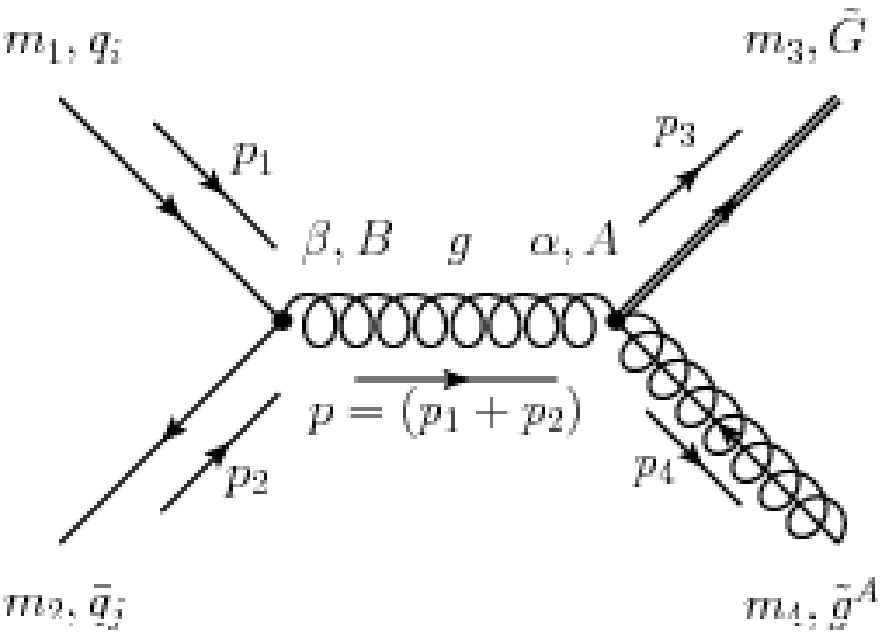}
&
\includegraphics[width=0.3\textwidth]{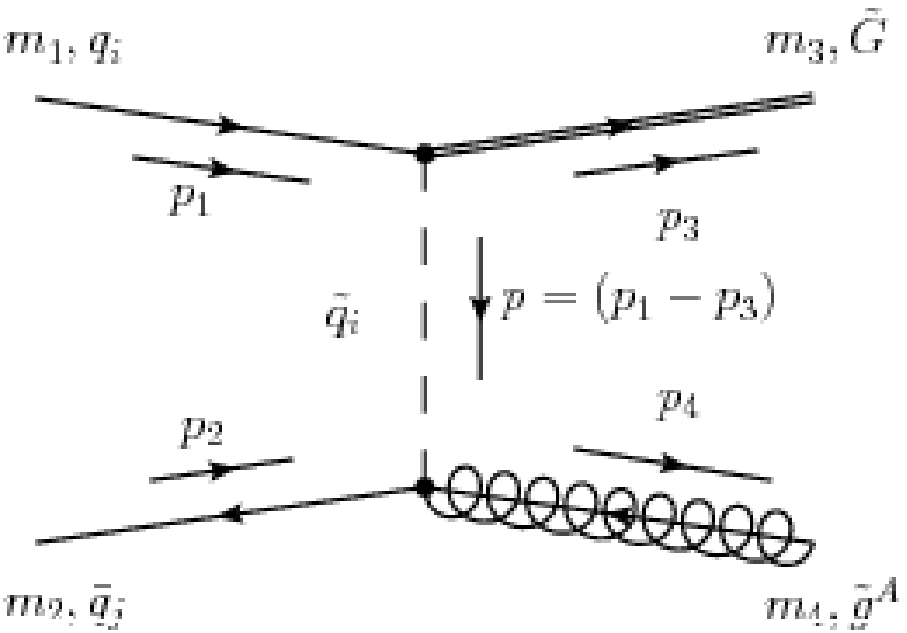}
&
\includegraphics[width=0.3\textwidth]{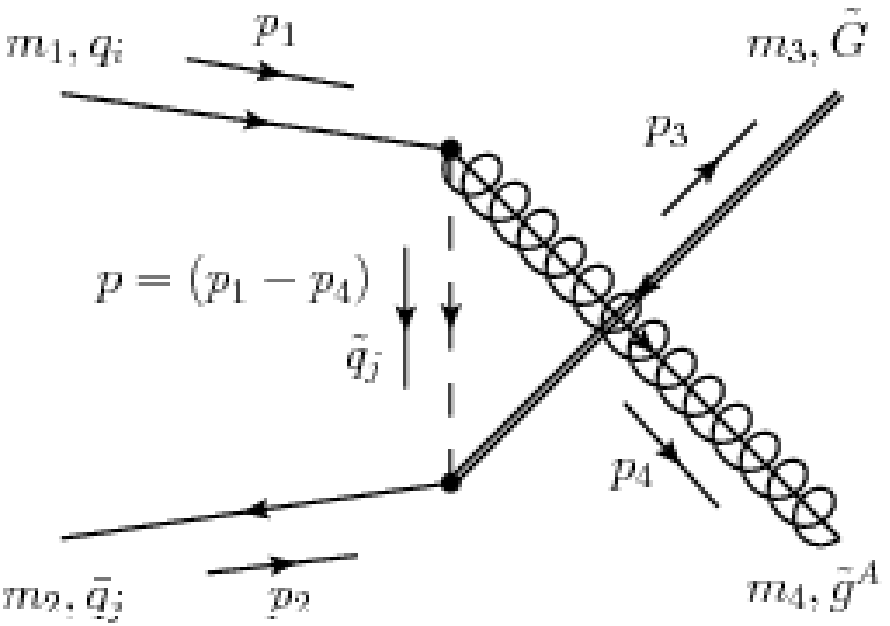} \\ [0.5cm]
(a) & (b) & (c) \\ 
\end{tabular} 
\end{center}
\caption{Tree level diagrams for the I process: $q_i + \bar{q}_j 
    \rightarrow \tilde{g}^A + \tilde{G}$ 
    scattering. 
}
\label{fig:i-diags}
\end{figure}

\begin{figure}[!htp]
\begin{center}
\begin{tabular}{ccc}
\includegraphics[width=0.3\textwidth]{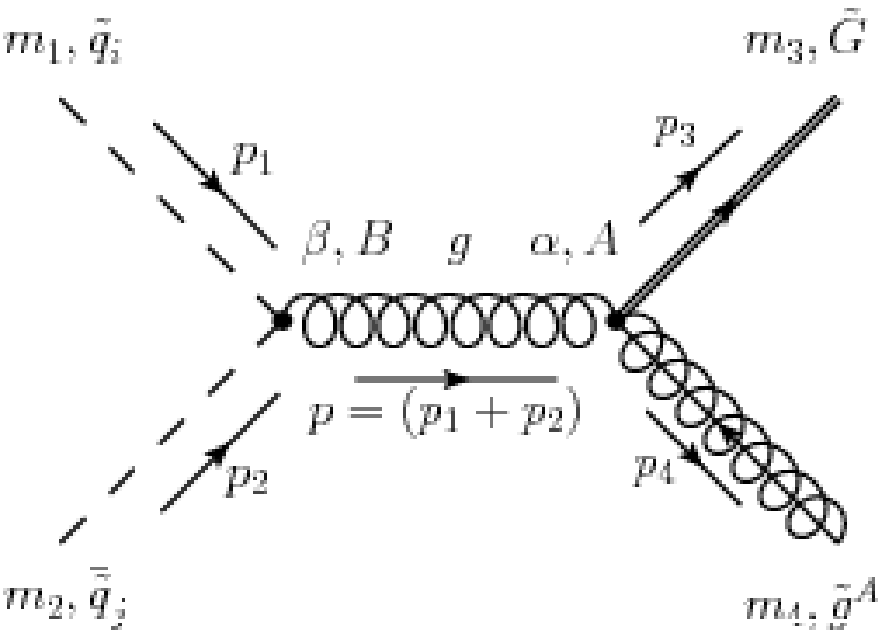}
&
\includegraphics[width=0.3\textwidth]{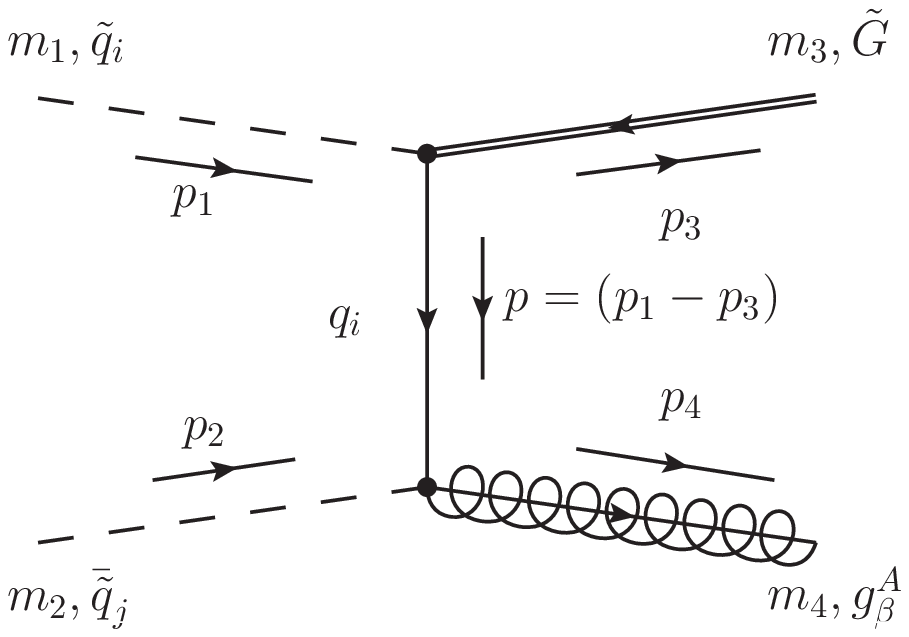}
&
\includegraphics[width=0.3\textwidth]{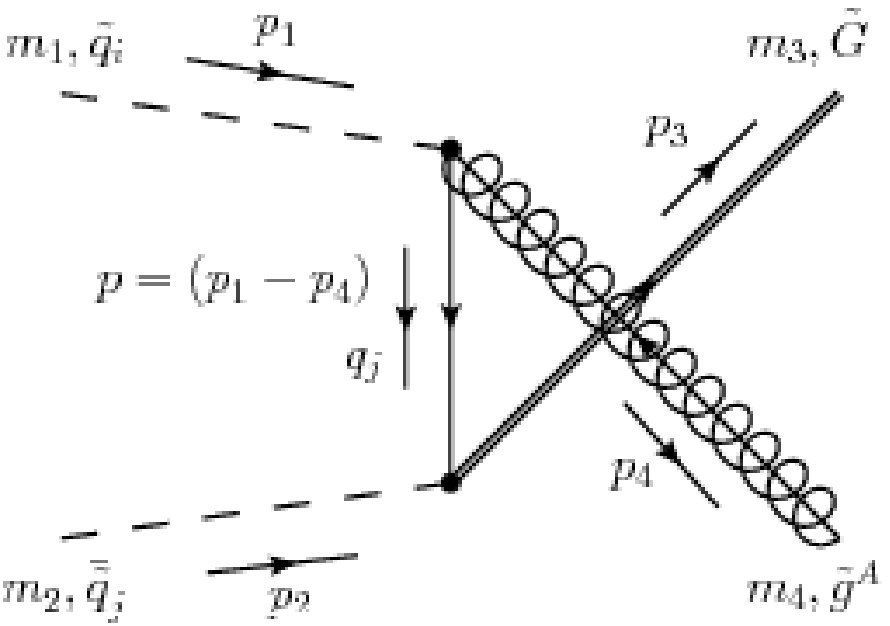} \\ [0.5cm]
(a) & (b) & (c) \\ 
\end{tabular} 
\end{center}
\caption{Tree level diagrams for the J process: 
    $\tilde{q}_i + \bar{\tilde{q}}_j 
    \rightarrow \tilde{g}^A + \tilde{G}$ scattering. 
}
\label{fig:j-diags}
\end{figure}

The production of gravitinos after inflation has been discussed in 
\cite{Nanopoulos:1983up,Krauss:1983ik,Falomkin:1984eu,Khlopov:1984pf,
Ellis:1984eq,Juszkiewicz:1985gg,Ellis:1984er,Kawasaki:1986my,
Khlopov:1993ye,Moroi:1993mb,Kawasaki:1994af,Bolz:2000fu,Cyburt:2002uv,
Giudice:1999am,Kawasaki:2004qu,Pradler:2006qh,Pradler:2006hh,
Rangarajan:2006xg,Rangarajan:2008zb,Rychkov:2007uq}.
The processes involving strong interactions denoted by A to J are 
\cite{Ellis:1984eq}
\begin{itemize}
\item A: $g^A + g^B \rightarrow \tilde{g}^C + \tilde{G}$
\item B: $g^A + \tilde{g}^B \rightarrow g^C + \tilde{G}$
\item C: $\tilde{q}_i + g^A \rightarrow q_j + \tilde{G}$,
$\bar{\tilde{q}}_i + g^A \rightarrow \bar{q}_j + \tilde{G}$
\item D: $q_i + g^A  \rightarrow \tilde{q}_j + \tilde{G}$,
$\bar{q}_i + g^A   \rightarrow \bar{\tilde{q}}_j + \tilde{G}$
\item E: $q_i + \bar{\tilde{q}}_j +  \rightarrow g^A + \tilde{G}$,
$\bar{q}_i + \tilde{q}_j   \rightarrow g^A + \tilde{G}$
\item F: $\tilde{g}^A + \tilde{g}^B \rightarrow \tilde{g}^C + \tilde{G}$
\item G: $q_i + \tilde{g}^A \rightarrow q_j + \tilde{G}$,
$\bar{q}_i + \tilde{g}^A \rightarrow \bar{q}_j + \tilde{G}$
\item H: $\tilde{q}_i + \tilde{g}^A \rightarrow \tilde{q}_j + \tilde{G}$,
$\bar{\tilde{q}}_i + \tilde{g}^A \rightarrow \bar{\tilde{q}}_j + \tilde{G}$
\item I: $q_i + \bar{q}_j \rightarrow \tilde{g}^A + \tilde{G}$
\item J: $\tilde{q}_i + \bar{\tilde{q}}_j \rightarrow \tilde{g}^A + \tilde{G}$
\end{itemize}
The above can be generalized for all gauge groups. We shall consider
$SU(3)_c$ only.  In the non-thermal scenario we only have to consider
processes involving the scattering of fermions and/or sfermions
\cite{Allahverdi:2005mz}. The reason is as follows: the gauge bosons
and gauginos have large masses induced by the flat direction vev. Due to
their large masses they decay quickly to lighter fermions and sfermions
and therefore do not participate in the production of gravitinos. The
processes not involving gauge bosons and gauginos as incoming particles
are the E, I and J processes.  These processes are shown in \figs
\ref{fig:e-diags}, \ref{fig:i-diags} and \ref{fig:j-diags}. (The second
E process is not shown and its contribution will be the same as that of
the first process.)
%
% \begin{widetext}
%

% \end{widetext}
%

In \refe \cite{Bolz:2000fu} it has been argued that the gravitino production
cross section should be proportional to 
\begin{equation}
%\frac{g_s^2}{M^2} \left( 1 + \frac{m^2_{\gt}}
\frac{g_s^2}{M^2} \left( 1 + \frac{m^2_{0}}
{3 m^2_{\widetilde{G}}} \right)\, ,
\label{eq:gtpropto}
\end{equation}
where $M = M_{Pl}/\sqrt{8\pi} = 2.4 \times 10^{18} \GeV$ is the reduced
Planck mass (and $m_\gt$ in the expression in
\refe \cite{Bolz:2000fu} has been replaced
by the supersymmetry breaking scale $m_0$).  
The factor $1$ is related to the contribution of the $\pm
3/2$ helicity states of the gravitino while the term proportional to
$m^2_{0}/(m^2_{\widetilde{G}})$ represents the contribution
%$m^2_{\tilde{g}}/(m^2_{\widetilde{G}})$ represents the contribution
of the helicity $\pm 1/2$ states.  To obtain the contribution of the
$\pm1/2$ helicity states, for $\sqrt{s}\gg m_\Gt$ we may express the
gravitino field as
\begin{equation}
\psi_\mu \rightarrow i\sqrt{\frac{2}{3}} \frac{1}{m_\Gt} \,\partial_\mu \psi,
\end{equation}
where $\psi$ represents a spin $1/2$ fermionic field. Cross sections
for spin $1/2$ gravitino production can then be obtained from the
corresponding effective lagrangian for $\psi$ \cite{Moroi:1995fs,
Drees:2004jm,Baer:2006rs}.  The total cross section is then obtained
in \refe \cite{Bolz:2000fu} using \eq \eqref{eq:gtpropto}.  

\section{Obtaining $\tkin$, $\tthr$, $t_G$ and $t_f$ and checking the 
feasibility of gravitino production}
\label{sec:discuss}

In this section we obtain $\tkin$ and $\tthr$ for the cases
$m_0 < \Gamma_d$ and $m_0 > \Gamma_d$ considered in \refe
\cite{Allahverdi:2005mz}.  Since gravitino production is suppressed by
lack of thermalization we impose the constraint that the $\Gamma_{kin,thr}
< H$ at $\td$ to study this effect.  We also impose the constraint
that the energy of the incoming quarks and squarks should be greater
than $3 m_{g,\gt}$ to avoid phase space suppression of the gravitino
production cross section by the $s-m_{g,\gt}^2$ factor in \eq
\eqref{eq:diffscatt}. These constraints give lower and upper bounds
on $\varphi_0$ respectively, where $\varphi_0$ is the flat direction
vev at $t_0 = m_0^{-1}$ when the flat direction starts oscillating.
In the  case of $m_0 < \Gamma_d$, these bounds are in fact incompatible
indicating that lack of thermalization effectively shuts off gravitino
production through phase space suppression. Surprisingly, we find that
even after thermalization, for both cases $m_0 \lessgtr \Gamma_d$,
gravitino production is phase space suppressed till the flat direction
condensate decays because of the large gluon and gluino mass.

We take the inflaton mass $m_\phi$ to be $10^{13} \GeV$ and $\alpha =
g_3^2/(4\pi)$ to be $5 \times 10^{-2}$ as applicable for the relevant
energy scales for $t \le \tthr$ and $\alpha = 0.1 $ for $t \gg \tthr$ as
relevant below.  We presume the inflaton decays at $\td = \Gamma_d^{-1}$
via $\phi \rightarrow X_1 + X_2$ so that the number density of the $X_1$
or $X_2$ particles equals that of the inflaton. The gluon mass is given by
\begin{equation}
m_g^2
= \alpha \varphi^2 =
\begin{cases}
    \alpha \varphi_0^2  & \textrm{for } t \leq t_0 \\
    \alpha \varphi_0^2 \paren{a_0 / a }^3 & \textrm{for } t \geq t_0 
\end{cases}
\end{equation}

\subsection{$m_0 < \Gamma_d$}
\label{ssec:tm0lGd}

For $m_0 < \Gamma_d$, the flat direction condensate starts
oscillating after the inflaton decays. For this case we take $m_0
= 100 \GeV$ and $\Gamma_d = 10^{4} \GeV$ as in \sect 3.1 of \refe
\cite{Allahverdi:2005mz}.  One must separately consider the time intervals
$t_d-t_0$, $t_0-\tkin$, $\tkin-\tthr$ and $t > \tthr$.

The gluon mass is given by $\sqrt{\alpha} \varphi$. The
energy of the incoming particles is $m_\phi (a_d/a)$ for $\td < t <
\tkin$. Since kinetic equilibration does not change the number of
particles while redistributing their energies, the average energy of
the incoming particles for $\tkin < t <\tthr$ is also given by $m_\phi
(a_d/a)$. Gravitino production for $t > \tthr$ is considered in 
\sect{\ref{ssec:YGtm0lGd}}.

\subsubsection{Non-equilibrium condition}

At $t = t_d$ we impose
\begin{equation}
H(t_d) = \Gamma_d > \Gamma_{kin} (t_d) = n \vev{\sigma v}_{kin} 
= \frac{\rho_\phi(t_d)}{m_\phi} 
    \frac{4\pi\alpha^2}{m_g^2}
\end{equation}
where we have used $\rho_\phi(t_d) = (3/ (8\pi)) M_{Pl}^2 \,\Gamma_d^2$
and $\vev{\sigma v}_{kin}\sim 4\pi\alpha^2/m_g^2$ 
\cite{Allahverdi:2005fq,Allahverdi:2005mz}. Then we get
\begin{align}
\Gamma_d &>
    \frac{3}{2} \frac{M_{Pl}^2 \Gamma_d^2} {m_\phi} 
    \frac{\alpha}{\varphi_0^2} 
\nonumber \\ \OR
\varphi_0 &>
    \sqrt{\frac{3}{2}} \left( \frac{\Gamma_d}{m_\phi} \right)^{1/2} 
    M_{Pl} \,\sqrt{\alpha}
\nonumber \\ \OR
\varphi_0 &>
    10^{14} \GeV , 
\label{eq:m0lGdlb}
\end{align}
where we have used $m_g^2 = \alpha \varphi_0^2$. $H(t_d) > \Gamma_{thr}(t_d)$
gives $\varphi_0 > 2 \times 10^{13} \GeV$.

Note that for $t_d < t < t_0$, $\Gamma_{kin}$ is given by
\begin{subequations}
\begin{align}
\Gamma_{kin} &= n \vev{\sigma v}_{kin}
\nonumber \\  
&= 
\frac{\rho_\phi(t_d)}{m_\phi} \left( \frac{a_d}{a} \right)^3
    \frac{4\pi\alpha^2}{m_g^2}
\label{eq:Gthr-m0lGd-td2t0-1}
\\ 
&\sim \frac{1}{a^3} \sim \frac{1}{t^{3/2}} ,
\end{align}
\end{subequations}
where we have used $n = n(t_d) (a_d/a)^3$.  Since $H \sim 1/t$, we see
that $\Gamma_{kin}$ is falling faster than $H$. So, if $H > \Gamma_{kin}$
holds at $t_d$, then it also holds for $t_d < t < t_0$.

\subsubsection{Equilibration times}
\label{sssec:eqtimes}

For $t > t_0$, $\Gamma_{kin}$ is given by 
\eq \eqref{eq:Gthr-m0lGd-td2t0-1} with $m_g^2 = \alpha \varphi_0^2 
(a_0/a)^3$. Therefore
\begin{subequations}
\begin{align}
\Gamma_{kin} 
&= \frac{3}{2} \frac{M_{Pl}^2 \Gamma_d^2}{m_\phi} \left( \frac{a_d}{a_0}
    \right)^3 \frac{\alpha}{\varphi_0^2} 
\label{eq:Gthr-m0lGd-t02tthr}
\\ 
&= \frac{3}{2} \frac{M_{Pl}^2 }{m_\phi \varphi_0^2}
    \, \Gamma_d^{1/2} \, m_0^{3/2} \alpha = \textrm{constant} .
\label{eq:gthr-m0lGd}
\end{align}
\end{subequations}
Since $H$ is decreasing with time, after some time $\tkin > t_0$, $H$
becomes less than $\Gamma_{kin}$, and later it becomes less than
$\Gamma_{thr}$. Setting $\tkin = \Gamma_{kin}^{-1}$
we get
\begin{subequations}
\begin{align}
\tkin &= \frac{2}{3} \frac{m_\phi}{M_{Pl}^2 \Gamma_d^{1/2}} 
\left(\frac{1}{m_0} \right)^{3/2} \frac{\varphi_0^2}{\alpha}
\label{eq:tthrm0lGd}
\\ \OR
\tkin &= 9 \times 10^{-30} \GeV^{-1}
\left(\frac{\varphi_0^2}{\GeV^2}\right).
\end{align}
\end{subequations}
$\Gamma_{thr}(t) = n \vev{\sigma v}_{thr}= \alpha \,\Gamma_{kin}$
as $\vev{\sigma v}_{thr}\sim 4\pi\alpha^3/m_g^2$ 
\cite{Allahverdi:2005fq,Allahverdi:2005mz}.
For $t>t_0$, $\Gamma_{thr}(t)\sim (1/a^3) \, (\alpha^3/m_g^2)
\sim$ constant, as for $\Gamma_{kin}$.  Then $\tthr = \alpha^{-1} \, \tkin$.

As we shall see below, there is also an upper bound on $\varphi_0$ which is
highly restrictive. Taking $\varphi_0 = 10^{14} \GeV$, we get,
\begin{equation}
\tkin = 9 \times 10^{-2} \GeV^{-1}.
\end{equation}
Note that $t_d = 10^{-4} \GeV^{-1}$ and $t_0 = 10^{-2} \GeV^{-1}$.
$\tthr = 2 \GeV^{-1}$.  For larger $\varphi_0$, both $\tkin$ and $\tthr$
will be larger.

\subsubsection{Avoiding phase space suppression}

Gravitinos are produced along with gluons or gluinos in the processes
discussed in \sect \ref{sec:revproc}.  
To avoid phase space suppression by the $s- m_{g,\gt}^2$
factor in the differential cross section we impose 
\begin{align}
3 \,m_{g,\tilde{g}} &< E_1 + E_2\,. 
\label{eq:pscond}
\end{align}
We apply this condition for gravitino production
before and after thermalization 
(taking $m_\gt\approx m_g$ in this analysis).
\\

%\paragraph{Before thermalization:}
\noindent
\textit{Before thermalization:}
We first consider gravitino production in the time interval $t_d
< t < t_0$.  
\Eq \eqref{eq:pscond} implies
\begin{align}
3 \sqrt{\alpha} \varphi_0 &< m_\phi \left( \frac{a_d}{a} \right).
\label{eq:ps-int1-m0lGd}
\end{align}
where $E_1$ and $E_2$ are the incoming energies in the CM frame.  We see
that the RHS falls like $1/a$, whereas the LHS is constant. We wish to
analyse whether delayed thermalization suppresses gravitino production.
Therefore to consider maximum gravitino production, we impose the above
inequality at $t_0$. If the above inequality holds at $t_0$, then it
will hold at earlier times till $t_d$ too.  Using the above inequality
we find an upper bound on $\varphi_0$.
\begin{align}
\varphi_0 &< \frac{m_\phi}{3 \sqrt{\alpha}} 
    \left( \frac{t_d}{t_0} \right)^{1/2}
\nonumber \\ \textrm{or,} \qquad
\varphi_0 &< \frac{m_\phi}{3 \sqrt{\alpha}} 
    \left( \frac{m_0}{\Gamma_d} \right)^{1/2}
\nonumber \\ \textrm{or,} \qquad 
\varphi_0 &< 1.5 \times 10^{12} \GeV. 
\label{eq:m0lGdub}
\end{align}
Thus we need $\varphi_0 < 1.5 \times 10^{12} \GeV$ to get gravitino
production from $t_d$ till time $t_0$.
For the time interval $t_0 < t < \tthr$, $m_{g,\gt} \sim \varphi \sim
1/a^{3/2}$ and the condition in \eq \eqref{eq:pscond} gives
\begin{align}
3 \,\sqrt{\alpha} \varphi_0 \left( \frac{a_0}{a} \right)^{3/2} 
    &< m_\phi \left( \frac{a_d}{a} \right).
\label{eq:ps-int2-m0lGd}
\end{align}
We see that the RHS falls like $1/a$ and the LHS falls like $1/a^{3/2}$. So
if the above inequality holds at $t_0$, which implies 
\eq \eqref{eq:m0lGdub} then it will hold for all time $t$ till $\tthr$.

We see that for the time interval $t_d < t < \tthr$ the upper bound on
$\varphi_0$ (\eq \eqref{eq:m0lGdub}) is smaller than the lower bound on
$\varphi_0$ (\eq \eqref{eq:m0lGdlb}). Thus, gravitino production for the
entire time interval $t_d < t < \tthr$ is not possible for the 
$m_0 < \Gamma_d$ case. 
\\

%\paragraph{After thermalization:}
\noindent
\textit{After thermalization:}
At $t=\tthr$, just after full equilibration, the average energy
of particles in the Universe falls which further inhibits gravitino
production.  Gravitino production can start at $t_G$ when $3 m_{g,\gt}$
becomes less than the incoming particle energy $(E_1 + E_2)$, i.e.,
\begin{align}
3 m_{g,\gt} &< E_1 + E_2
\nonumber \\ \OR \quad
    3 \sqrt{\alpha} \varphi_0 \paren{\frac{a_0}{a}}^{3/2} 
&< 
    2\times\frac{3}{2}  \,T_R \paren{\frac{\athr}{a}} \,,
\label{eq:TGcond}
\end{align}
where the subscript $R$ denotes reheating or full equilibration at $\tthr$. 
This implies
\begin{align}
    \sqrt{\alpha} \varphi_0 \paren{\frac{a_0}{\athr}}^{3/2}
    \paren{\frac{\athr}{a}}^{3/2} 
&<
    T_R \paren{\frac{\athr}{a}}
\nonumber \\ \OR \quad
    \sqrt{\alpha} \varphi_0 
    \paren{\frac{t_0}{\tthr}}^{3/4} 
&<
    T_R \paren{\frac{a}{\athr}}^{1/2}
\nonumber \\ \OR \quad
    \frac{\alpha \,\varphi_0^2}{T_R^2}
    \,\paren{m_0 \,\tthr}^{-3/2}
&<
    \paren{\frac{a}{\athr}}\,.
\label{eq:aGm0lGd}
\end{align}
Then,
\begin{align}
t_G &= \left[ \frac{\alpha \,\varphi_0^2}{T_R^2}
    \paren{\frac{1}{m_0}}^{3/2} \paren{\frac{1}{\tthr}} \right]^2
\nonumber \\ 
    &= 81 \paren{\frac{g_*}{\calA}}
    \frac{\alpha^2 \varphi_0^4 \,\Gamma_d^2 }{m_\phi^4 \,m_0^3}\,,
\label{eq:tGm0lGd}
\end{align}
where we have used \sect \ref{sssec:eqtimes} for $\tthr$ and the
expression for $T_R$ is derived later in \sect{\ref{ssec:TR}}.
The parameter $\calA$ that enters via $T_R$ is obtained to be $4 \times
10^{-5}$ in \sect{\ref{ssec:calA}}.

The perturbative flat direction decay rate is $\sim m_0^3 / \varphi^2$ 
\cite{Affleck:1984fy,Olive:2006uw} and the flat direction decays when
the decay rate is of order $H$ at a time $t_f$. Then
\begin{align}
t_f &= \frac{\varphi_f^2}{m_0^3}
\label{eq:tf}
\\
    &= \frac{\varphi_0^2}{m_0^3} \paren{\frac{a_0}{a_f}}^3
\nonumber \\
    &= \frac{\varphi_0^2}{m_0^3} \paren{\frac{t_0}{t_f}}^{3/2}
\nonumber \\
\OR t_f  &= \frac{\varphi_0^{4/5}}{m_0^{9/5}}\,. 
\label{eq:tfm0lGd}
\end{align}

We see that $t_G<t_f$ only for $\varphi_0 < 3 \times10^{12}\GeV$, which
conflicts with the lower bound on $\varphi_0$ in \eq \eqref{eq:m0lGdlb}.
Thus the non-equilibrium condition implies that the condensate decays
before gravitino production can commence in the radiation dominated era
after thermalization. Subsequent to the decay of the condensate there
will be gravtino production in the thermal Universe as discussed later
in \sect{\ref{ssec:YGtm0lGd}}.

\subsubsection{Relaxing the bound from phase space suppression}

If we relax the bound in \eq \eqref{eq:m0lGdub}, we may get
gravitino production for some part of the time interval $t_d < t <
\tthr$. We first consider $t_d < t < t_0$.  Using $a \sim t^{1/2}$ in
\eq \eqref{eq:ps-int1-m0lGd} and $t_d = \Gamma_d^{-1}$ we find gravitino
production is not phase space suppresed for
\begin{align}
t &< \frac{m_\phi^2}{\Gamma_d} \frac{1}{9 \,\alpha \,\varphi_0^2}
\nonumber \\ \OR
t &< 2 \times 10^{-6} \GeV^{-1} ,
\label{eq:tb-rlb-m0lGd}
\end{align}
for $\varphi_0 = 10^{14} \GeV$. Since this corresponds to times
less than $t_d = 10^{-4} \GeV^{-1}$, it implies that one cannot get
gravitino production anywhere in this time interval. 

We now consider $t_0 < t < t_{thr}$. Gravitino production can start at
some time after $t_0$, when the inequality in \eq \eqref{eq:ps-int2-m0lGd}
is satisfied. This is equivalent to
\begin{align}
3 \sqrt{\alpha} \varphi_0 \left( \frac{a_0}{a}\right)^{1/2}
&<
m_\phi \left( \frac{a_d}{a_0} \right) 
\nonumber \\ \OR
81 \,\alpha^2 \frac{\varphi_0^4}{m_0 \,m_\phi^4} \left( 
\frac{\Gamma_d}{m_0} \right)^2 
&< 
t \,\,.
\end{align}
For $\varphi_0 = 10^{14} \GeV$ this implies
\begin{equation}
t > 2 \times 10^{5} \GeV^{-1} \,.
\label{eq:tb-rub-m0lGd}
\end{equation}
Since this is larger than $\tthr = 2 \GeV^{-1}$, once again
there is no gravitino production in this time interval.

Therefore the conditions required for lack of kinetic and chemical
equilibration leads to a large outgoing gluon and gluino mass and thus
to phase space suppression that shuts off gravitino production for the
entire time interval from $t_d$ to $\tthr$ in the $m_0 < \Gamma_d$ case.
Increasing the value of $\varphi_0$ will not affect this conclusion as
this will only lower or raise the limits in \eqs \eqref{eq:tb-rlb-m0lGd}
and \eqref{eq:tb-rub-m0lGd} respectively. Similarly, gravitino production
is phase space suppressed in the radiation dominated era before the
condensate decays at $t_f$ and increasing the value of $\varphi_0$
will not alter this conclusion.  (Decreasing the value of $\varphi_0$
to avoid phase space suppression will lead to the standard scenario of
reheating and gravitino production in a thermal Universe.)

Thus, for the parameters chosen above for the scenario under
consideration, there is no gravitino production for the case $m_0 <
\Gamma_d$ before the condensate decays.

\subsection{$m_0 > \Gamma_d$ }
\label{ssec:tm0gGd}

We now consider the case $m_0 > \Gamma_d$, where the flat direction
condensate starts oscillating before the inflaton decays. $m_0
\sim 100$ GeV and we take $\Gamma_d = 10$ GeV as in \sect{3.1} of 
\refe \cite{Allahverdi:2005mz}. 

\subsubsection{Non-equilibrium condition and relevant time scales}
\label{sssec:noeqcond-m0gGd}

The absence of kinetic equilibration of the inflaton decay
products requires $\Gamma_{kin} < H$. $\Gamma_{kin}$ is given by \eq
\eqref{eq:Gthr-m0lGd-t02tthr}.  We require $\Gamma_{kin}$ to be much
less than $H$ at $t_d$ to obtain a significant phase of non-equilibrium
since $H$ decreases with time while $\Gamma_{kin}$ remains constant.
Then applying the non-equilibrium condition at $t_d$, and setting
$\varphi_d^2 = \varphi_0^2 (a_0^3/a_d^3)$  and $H(t_d) = \Gamma_d$, we get
\begin{align}
\Gamma_d &\gg \Gamma_{kin} (t_d)  
\nonumber \\ \textrm{or,}\qquad\qquad
\varphi_0^2
    &\gg
    \frac{3}{2 m_\phi} M_{Pl}^2 \,\Gamma_d \,\alpha
    \left(\frac{a_d}{a_0}\right)^3 
\nonumber \\ \textrm{or,}\qquad\qquad
\varphi_0^2
    &\gg 
    \frac{3}{2 m_\phi} M_{Pl}^2 \,\Gamma_d \,\alpha 
    \left(\frac{t_d}{t_0}\right)^2 
\nonumber \\ \textrm{or,}\qquad\qquad
\varphi_0^2
    &\gg 
    \frac{3}{2 m_\phi} M_{Pl}^2 \,\Gamma_d \,\alpha 
    \left(\frac{m_0}{\Gamma_d}\right)^2
\nonumber \\ \textrm{or,} \qquad\qquad
\varphi_0 &> 3 \times 10^{13} \GeV \,, 
\label{eq:LBvarphi}
\end{align}
where we have used $a \sim t^{2/3}$ between $t_0$ and $t_d$ as relevant
for the inflaton oscillating in a quadratic potential before decay.
Chemical non-equilibration at $t_d$ requires $\varphi_0 > 3 \times 10^{13}
\GeV \sqrt{\alpha} \simeq 7 \times 10^{12} \GeV$. 

Thus, for $\varphi_0 < 7 \times 10^{12} \GeV$ the Universe is in
kinetic and chemical equilibrium at $t_d$. For $ 7 \times 10^{12} \GeV < 
\varphi_0 < 3 \times 10^{13} \GeV$ the Universe is in kinetic but not in 
chemical equilibrium at $t_d$. Finally, for $\varphi_0 > 3 \times 10^{13} \GeV$
the Universe is neither in kinetic nor chemical equilibrium at $t_d$.

To obtain $\tkin$ we use \eq \eqref{eq:Gthr-m0lGd-t02tthr} and $a \sim
t^{2/3}$ for $t_0 < t < \td$. Then 
\begin{align}
\tkin^{-1} = \Gamma_{kin} &=
    \frac{3}{2} \frac{\alpha \,M_{Pl}^2 \,\Gamma_d^2}{m_\phi \,\varphi_0^2}
    \paren{\frac{t_d}{t_0}}^2
\nonumber \\ &= 
    \frac{3}{2} \frac{\alpha \,M_{Pl}^2 \,\Gamma_d^2}{m_\phi \,\varphi_0^2}
    \paren{\frac{m_0}{\Gamma_d}}^2
\nonumber \\ &= 
    \frac{3}{2} \frac{\alpha \,M_{Pl}^2 \,m_0^2}{m_\phi \,\varphi_0^2}.
\label{eq:Gtthrm0gGd}
\end{align}
As before, $\Gamma_{thr} = \alpha \Gamma_{kin}$ and 
$\tthr = \Gamma_{thr}^{-1} = \alpha^{-1} \tkin$.

\subsubsection{Avoiding phase space suppression}

For gravitino production not to be phase space suppressed by the
$s-m_{g,\tilde{g}}^2$ factor for $t_d<t<\tthr$ we again impose the
condition of \eq \eqref{eq:pscond} and get \eq \eqref{eq:ps-int2-m0lGd}.
To consider maximum gravitino production we assume that this condition
is valid from $t_d$.  Using \eq{\eqref{eq:ps-int2-m0lGd}} at $\td$ gives
\begin{equation}
3 \sqrt{\alpha} \varphi_d  < m_\phi,
\end{equation}
where $\varphi_d = \varphi_0 (a_0/a_d)^{3/2} = \varphi_0 (t_0/t_d) =
\varphi_0 (\Gamma_d/m_0)$.  Therefore we get
\begin{equation}
\varphi_0 < 1.5 \times 10^{14} \GeV.
\label{eq:UBvarphi}
\end{equation}
Since the LHS of \eq{\eqref{eq:ps-int2-m0lGd}} falls faster than the RHS,
this upper bound on $\varphi_0$ also ensures the viability of gravitino
production at  times after $\td$.

We shall take $\varphi_0 = 10^{14} \GeV$ hereafter so that gravitino
production commences at $t_d$ and there is neither kinetic nor chemical
equilibration at $t_d$. For this value of $\varphi_0$, $\tkin = 0.9 
\GeV^{-1}$ and $\tthr = 20 \GeV^{-1}$.

For gravitino production after $\tthr$, \eq{\eqref{eq:TGcond}} implies
\begin{align}
    3 \sqrt{\alpha} \varphi_0 
    \paren{\frac{a_0}{a_d}}^{3/2} 
    \paren{\frac{a_d}{\athr}}^{3/2}
    \paren{\frac{\athr}{a}}^{3/2}
&< 
    2\times{\frac{3}{2}} \,T_R \paren{\frac{\athr}{a}}   
\nonumber \\ \OR \quad
     \sqrt{\alpha} \varphi_0 
    \paren{\frac{t_0}{t_d}}
    \paren{\frac{t_d}{\tthr}}^{3/4}
&< 
     T_R \paren{\frac{a}{\athr}}^{1/2}   
\nonumber \\ \OR \quad
    \frac{\alpha \,\varphi_0^2}{T_R^2} 
    \paren{\frac{\Gamma_d}{m_0}}^2
    \paren{\frac{1}{\Gamma_d \,\tthr}}^{3/2}
&< 
    \paren{\frac{a}{\athr}} \,.
\label{eq:aGm0gGd}
\end{align}
Then
\begin{align}
t_G  
    &= \left[ \frac{\alpha \,\varphi_0^2}{T_R^2} 
    \paren{\frac{\Gamma_d^{1/2}}{m_0^2}}
    \frac{1}{\tthr} \right]^2 
\nonumber \\
    &= 81\paren{\frac{g_*}{\calA}}
    \frac{\alpha^2 \varphi_0^4 \,\Gamma_d^3 }{m_\phi^4 \,m_0^4}\,,
\label{eq:tGm0gGd}
\end{align}
where we have used the expression for $\tthr$ from \sect
\ref{sssec:noeqcond-m0gGd} and for $T_R$ as in \eq \eqref{eq:TRcommon}
or \eqref{eq:TRm0gGd}.  $\calA = 4\times 10^{-11}$ as obtained in
\sect{\ref{ssec:calA}}.

The time when the condensate decays, $t_f$, is given by \eq{\eqref{eq:tf}}.
Then
\begin{align}
t_f &= \frac{\varphi_0^2}{m_0^3} \paren{\frac{a_0}{a_d}}^3
    \paren{\frac{a_d}{a_f}}^3 
\nonumber \\
    &=  \frac{\varphi_0^2}{m_0^3} \paren{\frac{t_0}{t_d}}^{2}
    \paren{\frac{t_d}{t_f}}^{3/2} 
\nonumber \\
    &= \frac{\varphi_0^2}{m_0^3} \paren{\frac{\Gamma_d}{m_0}}^{2}
    \paren{\frac{1}{\Gamma_d t_f}}^{3/2}
\nonumber \\
\OR t_f &= \frac{\varphi_0^{4/5} \Gamma_d^{1/5}}{m_0^2} \,.
\label{eq:tfm0gGd}
\end{align}

For $\varphi_0 = 10^{14} \GeV$, $t_G$ is $5 \times 10^{11} \GeV^{-1}$
while $t_f$ is $3 \times 10^{7} \GeV^{-1}$. Thus, there is no gravitino
production in the radiation dominated era after thermalization before
the flat direction condensate decays.

\subsubsection{Relaxing the bounds}

The lower and upper bounds on $\varphi_0$ in \eqs \eqref{eq:LBvarphi}
and \eqref{eq:UBvarphi} are restrictive. For later use we relax the upper
bound from phase space suppression. If $\varphi_0 > 1.5 \times 10^{14} \GeV$
it implies that gravitino production commences at a time $t_p > t_d$.
To get an expression for $t_p$ we start with \eq{\eqref{eq:ps-int2-m0lGd}}
and solve for $a_p$, the scale factor at which gravitino
production starts. We thus get
\begin{align}
m_\phi \frac{a_d}{a_p} &=
3\sqrt{\alpha} \varphi_0 \left(\frac{a_0}{a_p}\right)^{3/2} 
\nonumber \\
\textrm{or,} \qquad
m_\phi \frac{a_d}{a_p} &= 
3 \sqrt{\alpha} \varphi_0 \left(\frac{a_0}{a_d}\right)^{3/2} 
 \left(\frac{a_d}{a_p}\right)^{3/2}
\nonumber \\
\textrm{or,} \qquad t_p &= \frac{1}{\Gamma_d} \left[ 3 \sqrt{\alpha} 
\varphi_0 \left( \frac{\Gamma_d}{m_0} \right) \frac{1}{m_\phi} \right]^4,
\label{eq:tpm0gtGd}
\end{align}
where we have used the fact that for $t_0 < t < t_d$ the scale factor
$a \sim t^{2/3}$ and for $t_d < t < t_p$, $a \sim t^{1/2}$.  In this case one
will get an extra factor of $1/(\Gamma_d \,t_p)^{7/2}$ along with $t_d$
replaced by $t_p$ in the expressions for gravitino number density
generated before kinetic equilibration in \eqs{\eqref{eq:nGtEproc}},
\eqref{eq:nGtIproc} and \eqref{eq:nGtJproc}.

We remark in passing that if $7 \times 10^{12} \GeV < \varphi_0 < 3
\times 10^{13} \GeV$ and one has kinetic equilibration of the inflaton 
decay products at $t_d$ without chemical equilibration, then one would 
obtain gravitino production using only \sect{\ref{ssec:tkin2tthr}}, with 
$\Gamma_{kin}$ set equal to $\Gamma_d$.

\section{Calculations of $\sigma_{CM}$}
\label{sec:sigmaCM}

Having discussed the feasibility of gravitino production in a
non-thermal Universe in the context of phase space availability we now
calculate the cross section of gravitino production. The expressions
for $|\mathcal{M}|^2$ in this section include an average over spins and
colors of the initial and final state particles but are for a specific
flavor for quarks and squarks. The cross sections obtained below are for the
dominant $\pm 1/2$ helicity states of the gravitino and will be used for
obtaining the collision integral (the $A$ terms) for $t_d < t < \tkin$, as
in \ref{sec:Aterms}. 

\subsection{Differential scattering cross sections}

The differential scattering cross sections in the CM frame is as given
in \eq{\eqref{eq:diffscatt}}. 
We use $|\mathcal{M}|^2$ as provided in Table 1 of \refe \cite{Bolz:2000fu}.
Then
\begin{subequations}
\begin{align}
|\mathcal{M}_E|^2 &=
\frac{g_s^2}{M^2} \,\frac{ \sum_{A,i,j} \,|T^A_{ji}|^2}
{\,g_1\,g_2\,g_3\,g_4} 
\left( 1 + \frac{m_0^2}{3 m_\Gt^2} \right)
\, (-2t)
\label{eq:mE2}
\\
|\mathcal{M}_I|^2 &=
\frac{g_s^2}{M^2} \,\frac{ \sum_{A,i,j} \,|T^A_{ji}|^2}
{g_1\,g_2\,g_3\,g_4}
\left( 1 + \frac{m_0^2}{3 m_\Gt^2} \right)
    \,\frac{(-8 \,t) \,(s+t)}{s}
\label{eq:mI2}
\\
|\mathcal{M}_J|^2 &=
\frac{g_s^2}{M^2} \frac{ \sum_{A,i,j} |T^A_{ji}|^2}
{g_1\,g_2\,g_3\,g_4} 
\left( 1 + \frac{m_0^2}{3 m_\Gt^2} \right)
    \,\frac{2 \,(s^2 + 2 ts + 2 t^2) }{s} \,,
\label{eq:mJ2}
\end{align}
\end{subequations}
where $T^A$ is the generator of the $SU(3)$ group in the fundamental
representation, and $i,j$ represent the color of the incoming particles.
$|\mathcal{M}_I|^2$ above differs from \refe \cite{Bolz:2000fu}
by a factor of 2 because we sum over both chiralities of the quarks.
The expressions in \Tab 1 of \refe \cite{Bolz:2000fu} do not include
an average over incoming and final states and hence we have a sum
over $A,i,j$ and an additional factor of $g_1 \,g_2\,g_3\,g_4$ in the
denominators of $|\mathcal{M}_{E,I,J}|^2$.

Using  \eq\eqref{eq:diffscatt} in the limit that the incoming energy
$\sqrt{s}$ is much larger than all masses, the differential scattering
cross sections for the three processes are
\begin{subequations}
\begin{align}
\frac{d\sigma_E}{d\Omega} &= -\frac{ \sum_{A,i,j} g_s^2
\,|T^A_{ji}|^2}
{32 \,g_1\,g_2 \,M^2 \,\pi ^2 s} 
\left( 1 + \frac{m_0^2}{3 m_\Gt^2} \right)
t 
\\
\frac{d\sigma_I}{d\Omega} &= -\frac{ \sum_{A,i,j} g_s^2
 \,|T^A_{ji}|^2}
 {8 \,g_1\,g_2 \,M^2 \,\pi ^2 s^2}
\left( 1 + \frac{m_0^2}{3 m_\Gt^2} \right)
 t(s+t)
\\
\frac{d\sigma_J}{d\Omega} &= 
%- 
\frac{ \sum_{A,i,j} g_s^2 
\,|T^A_{ji}|^2}
{32 \,g_1\,g_2 \, M^2 \,\pi ^2 s^2 }
\left( 1 + \frac{m_0^2}{3 m_\Gt^2} \right)
(s^2 + 2ts + 2t^2) \,.
\end{align}
\end{subequations}

\subsection{Scattering cross sections}

The cross section $\sigma_{CM}$ is obtained by integrating the differential
cross section.  The scattering cross sections for the E, I and J processes are 
\begin{subequations}
\begin{align}
\sigma_E &= \frac{ \sum_{A,i,j} g_s^2 
\,|T^A_{ji}|^2 }
{16\,g_1\,g_2 \,M^2 \,\pi  } 
\left( 1 + \frac{m_0^2}{3 m_\Gt^2} \right)
\label{eq:sigmae}
\end{align}
\begin{align}
\sigma_I &= \frac{ \sum_{A,i,j} g_s^2 
\,|T^A_{ji}|^2} {
12\,g_1\,g_2 \,M^2 \,\pi  } 
\left( 1 + \frac{m_0^2}{3 m_\Gt^2} \right)
\label{eq:sigmai}
\end{align}
\begin{align}
\sigma_J &= \frac{ \sum_{A,i,j} g_s^2 
\,|T^A_{ji}|^2}{
12 \,g_1\,g_2 \,M^2 \,\pi   } 
\left( 1 + \frac{m_0^2}{3 m_\Gt^2} \right)\,.
\label{eq:sigmaj}
\end{align}
\end{subequations}

\section{Gravitino abundances}
\label{sec:Gtabund}

Having obtained the cross section $\sigma_{CM}$ we use the procedure 
given in \sect{\ref{sec:boltzmann}} to calculate the number density of
gravitinos via the E, I and J processes using the integrated Boltzmann 
equation. For $m_0 < \Gamma_d$ we only consider $t>\tthr$. For
$m_0 > \Gamma_d$, we calculate the number density separately from $t_d$
to $\tkin$, $\tkin$ to $\tthr$ and for $t> \tthr$.  Parameters needed
to evaluate the gravitino abundance are given in \sect{\ref{sec:params}}, 
\ref{sec:bparam} and \ref{sec:mgt}.

\subsection{$m_0 < \Gamma_d$}
\label{ssec:YGtm0lGd}

As shown in \sect{\ref{ssec:tm0lGd}} there is no gravitino production
until the flat direction condensate decays at $t_f$. The temperature
$T_f$ at $t_f$ is derived in \sect{\ref{sec:Tf}} and equals $1 \times
10^5 \GeV$. The gravitino abundance generated after $t_f$ is given by the
standard expression for gravitino production in a thermal Universe after
inflation with the reheat temperature replaced by $T_f$. For this low
value of $T_f$ the gravitino abundance will be small enough to satisfy
cosmological constraints.

\subsection{$m_0 > \Gamma_d$}
\label{ssec:YGtm0gGd}

\subsubsection{$t_d$ to $t_{kin}$}
\label{sssec:td2tkin}

The $A$ terms for the processes E, I, J that enter the RHS
of the integrated Boltzmann \eq \eqref{eq:mbltzeq} for the
phase space distribution in \eq{\eqref{eq:fi1}} are given in 
\ref{sec:Aterms}. 

For the process E,
\begin{align}
\frac{dn_{\widetilde{G}}^{(E)}}{dt} + 3 H n_{\widetilde{G}} = A_E 
&=
144 \,
B \,
\,m_\phi^4 \left( \frac{a_d}{a} \right)^6
\nonumber \\ &=
144\,
B \,
\, m_\phi^4 \left( \frac{t_d}{t} \right)^3
\nonumber \\ &=
144 \,Q \,t^{-3} \, ,
\label{eq:bernE}
\end{align}
where
\begin{align}
B &= \frac{ 3 \sum_{A,i,j} g_s^2 \,M_{Pl}^2 \,|T^A_{ji}|^2\,\Gamma_d^4}
    {256 \,\pi^2 \,g_1\,g_2 \,m_\phi^6}
   \left( 1 + \frac{m_0^2}{3 m_\Gt^2}\right) 
\\ 
Q &= \frac{B m_\phi^4}{\Gamma_d^3}
\label{eq:qe}
\end{align}
The parameter $B$ is derived in \ref{sec:bparam}.  
Solving \eq{\eqref{eq:bernE}} we then get
\begin{subequations}
\begin{align}
n_\Gt^{(E)}(t) &= 
144
K \left[ 1 - \paren{\frac{t_d}{t}}^{1/2} \right]
\paren{\frac{t_d}{t}}^{3/2}
\label{eq:nGtEproc}
\\ &=
    144
    K \left[ 1 - \paren{\frac{a_d}{a}} \right]
\paren{\frac{a_d}{a}}^{3},
\end{align} 
\end{subequations}
where $K$ is defined as
\begin{equation}
K = \frac{2Bm_\phi^4 }{\Gamma_d}
\label{eq:Kparam}
\end{equation}
Similarly for the process I,
\begin{align}
\frac{dn_{\widetilde{G}}^{(I)}}{dt} + 3 H n_{\widetilde{G}} = A_I 
&=  
48\, B \,  
\, m_\phi^4 \left( \frac{a_d}{a} \right)^6
\nonumber \\ &=
    48\,
    B \,
    \, m_\phi^4 \left( \frac{t_d}{t} \right)^3
\nonumber \\ &= 
    48 \, Q \,t^{-3} \, .
\label{eq:bernI}
\end{align}
\Eq \eqref{eq:bernI} implies that
\begin{subequations}
\begin{align}
n_{\widetilde{G}}^{(I)}(t) &=
48
K \left[ 1 - \paren{\frac{t_d}{t}}^{1/2} \right]
\paren{\frac{t_d}{t}}^{3/2}
\label{eq:nGtIproc}
\\ &=
    48
    K \left[ 1 - \paren{\frac{a_d}{a}} \right]
\paren{\frac{a_d}{a}}^{3}\,.
\end{align}
\end{subequations}
For the process J,
\begin{align}
\frac{dn_{\widetilde{G}}^{(J)}}{dt} + 3 H n_{\widetilde{G}} = A_J 
&=
96\,
B \,
\, m_\phi^4 \left( \frac{a_d}{a} \right)^6
\nonumber \\ &=
96\,
B \,
\, m_\phi^4 \left( \frac{t_d}{t} \right)^3
\nonumber \\ &= 
96 \, Q \,t^{-3} \,.
\label{eq:bernJ}
\end{align}
\Eq \eqref{eq:bernJ} implies that
\begin{subequations}
\begin{align}
n_{\widetilde{G}}^{(J)}(t) &= 
96
K \left[ 1 - \paren{\frac{t_d}{t}}^{1/2} \right]
\paren{\frac{t_d}{t}}^{3/2}
\label{eq:nGtJproc}
\\ &=
    96
    K \left[ 1 - \paren{\frac{a_d}{a}} \right]
\paren{\frac{a_d}{a}}^{3}.
\end{align}
\end{subequations}

One observes that the factor in the square brackets in the above
expressions is related to gravitino production while the factor $\sim
1/a^3$ reflects dilution due to Hubble expansion. Because of the fast
decline with time of the source term on the RHS of the integrated
Boltzmann equation, the production of gravitinos shuts off quickly and
then the number density falls as $1/a^3$.

\subsubsection{$\tkin$ to $\tthr$}
\label{ssec:tkin2tthr}

For $\tkin$ to $\tthr$ we evaluate A defined in \eq{\eqref{eq:mbltzeq}} with
the phase space distribution function given in \eq{\eqref{eq:fi2}}. 
\begin{align}
A &= \exp(\xi_1/T) \exp(\xi_2/T) \int ~d\Pi_1 ~d\Pi_2 
\exp(-E_1/T) \exp(-E_2/T) W_{12}
\nonumber \\
&= \exp(\xi_1/T) \exp(\xi_2/T) ~n_1^{eq} n_2^{eq} \left[
\frac{1}{n_1^{eq} n_2^{eq}} \int ~d\Pi_1 ~d\Pi_2 
\exp(-E_1/T) \exp(-E_2/T) W_{12} \right ]
\nonumber \\
&\equiv n_1(t) n_2(t) \vev{\sigma v} , 
\label{eq:A2} 
\end{align}
where $\vev{\sigma v}$ is the expression in square brackets and 
\begin{align}
n_i &= \frac{g_i}{(2\pi)^3} \int \exp[-(E_i -\xi_i)/T] ~d^3p_i 
\qquad \textrm{ for } i = 1,2
\nonumber \\
&= \calA_i ~n_i^{eq} \nonumber \\
&= \calA_i ~\frac{g_i}{\pi^2} T^3 .
\end{align}
Here,
\begin{align}
\calA_i \equiv \exp(\xi_i/T) &< 1 
\qquad \textrm{ for } \tkin < t < \tthr,
\nonumber \\
&= 1 \qquad \textrm{ for } t = \tthr,
\end{align}
and
\begin{equation}
n_i^{eq} = \frac{g_i}{(2\pi)^3} \int \exp(-E_i/T) \,d^3p_i \,, 
\end{equation}
and we have ignored factors of $\mathcal{O}(1)$.  
Our expression for $n_i$ includes $g_i$, unlike in \refe
\cite{Allahverdi:2005mz}.  The number density of all particles in the
Universe is
\begin{align}
n = \sum_j n_j &= \sum_j \calA_j \frac{g_j}{\pi^2} T^3 
\equiv \calA ~\frac{T^3}{\pi^2} \,. 
\end{align}
$\calA \rightarrow g_*$ in thermal equilibrium, where $g_*$ is the
total number of relativistic degrees of freedom for a fully thermalized
plasma. For the minimal supersymmetric Standard Model (MSSM) $g_* =
228.75$.  Since we can express $\calA$ for $\tkin$ to $\tthr$ in terms
of the standard thermally averaged cross section we can therefore use
expressions for $\vev{\sigma v}$ from the literature.

In kinetic equilibrium, temperature is well defined and we choose to
re-express the integrated Boltzmann \eq \eqref{eq:mbltzeq}
\begin{equation}
\dot{n}_{\Gt} + 3 H n_{\Gt} = \vev{\sigma v} n_1 n_2 ,
\end{equation}
as 
\begin{equation}
\dot{T} \frac{dn_{\Gt}}{dT} + 3 H n_{\Gt} = \vev{\sigma v} n_1 n_2 ,
\end{equation}
where a sum over all processes E, I, J is implicit. Defining $Y' = n/s'$,
where
$s' = \frac{2\pi^2}{45} \calA \,T^3$ , 
we can rewrite the Boltzmann equation as
\begin{equation}
\dot{T} \frac{dY'}{dT} = \vev{\sigma v} Y_1' n_2 .
\end{equation}
Since $T \sim 1/a$, $\dot{T}/T = - H$ and so 
\begin{equation}
\dot{T} = - \left( \frac{\calA \pi^2}{90} \right)^{1/2} \frac{T^3}{M} ,
\end{equation}
where we have used $H^2 = \rho/(3 M^2)$ and $\rho = (\pi^2/30) \calA T^4$.
Expressions for $\vev{\sigma v}$ from \refe \cite{Ellis:1984eq}, were
used in \refe \cite{Allahverdi:2005mz}.  These results were for the
$\pm 3/2$ helicity states of the gravitino. Including the $\pm 1/2$
helicity states we get
\begin{subequations}
\begin{align}
\vev{\sigma v}_E &= 
\frac{1}{32 M^2} \, \paren{1 + \frac{m_0^2}{3m_\Gt^2}}
    \,48 \alpha
\\
\vev{\sigma v}_I &= \frac{1}{32 M^2} \,\paren{1 + \frac{m_0^2}{3m_\Gt^2} }
    \,16 \alpha
\\
\vev{\sigma v}_J &= \frac{1}{32 M^2} \, \paren{1 + \frac{m_0^2}{3m_\Gt^2}}
    \,8 \alpha \,,
\end{align}
\label{eq:sigmav}
\end{subequations}
where we have considered only the $SU(3)_c$ gauge group.  The gravitino
abundance just before full thermalization at $\tthr$ is
\begin{align}
Y_\Gt'(\tthr) = 
\frac{n_\Gt(\tthr)}{s'(\tthr)}=
Y_\Gt'(\tkin)  
    + \frac{45}{2 \pi^6} \paren{\frac{90}{\mathcal{A}\pi^2}}^{1/2}
    \frac{1}{\mathcal{A}}\,
    \int_{T_{min}}^{T_{kin}} \vev{\sigma v} \calA_1 g_1 
    \,\calA_2 g_2 \,M dT 
\end{align}
where  $s'(\tthr)$ and $T_{min}$ are the entropy density and temperature
at $\tthr$ just before thermal equilibration. $s'(\tthr) = (2\pi^2/45)
\calA T_{min}^3$ and $\vev{\sigma v}$ includes a sum over all the
processes in \eq{\eqref{eq:sigmav}}.  $Y_\Gt'(\tkin)$ is obtained using
\eqs{\eqref{eq:nGtEproc}}, \eqref{eq:nGtIproc} and \eqref{eq:nGtJproc}
and the temperature at $\tkin$ needed for the entropy is given by \eq
\eqref{eq:Tkin}. Then
\begin{align}
Y_\Gt'(\tkin) &= \frac{n_\Gt(t_{kin})}{s(t_{kin})}
\nonumber \\[0.2cm] &=
\frac{\sum_l q_l \,K \left[ 1 - \paren{t_d / t_{kin}}^{1/2} \right]
    \paren{t_d / t_{kin}}^{3/2}}
    {\paren{2\pi^2 /45}\calA \,T_{kin}^3}
\nonumber \\[0.2cm] &=
\frac{\sum_l q_l \,K \left[ 1 - \paren{\Gamma_{kin} / \Gamma_d }^{1/2} 
    \right] \paren{\Gamma_{kin} /\Gamma_d }^{3/2}}
    {\paren{2\pi^2 /45 }\calA \,\paren{m_\phi /3 }^3
    \paren{\Gamma_{kin} / \Gamma_d }^{3/2}}
\nonumber \\[0.2cm] &=
\frac{\sum_l q_l \,K \left[ 1 - \paren{ \Gamma_{kin} / \Gamma_d }^{1/2}
    \right]} {\paren{2\pi^2 /45 }\calA \,\paren{ m_\phi /3 }^3} \,,
\end{align}
where $q_l = 144, 48, 96$ for the $l = E, I, J$ processes respectively.
Note that $K\propto B$ is also process dependent, 
as mentioned in \ref{sec:bparam}.

After equilibration at $\tthr$ the temperature falls to
\begin{equation}
T_R = \left( \frac{\calA}{g_*} \right)^{1/4} T_{min} ,
\label{eq:TrTmin}
\end{equation}
Then, 
\begin{align}
Y_{\Gt}(\tthr) 
&= Y'_{\Gt}(\tthr) \frac{s'(\tthr)}{s(\tthr)} 
\nonumber \\ &= 
    \left(\frac{\calA}{g_*} \right)^{1/4} Y'_{\Gt}(\tkin) 
    + \frac{0.07}{g_*^{3/2}}   
    \left(\frac{g_*}{\calA} \right)^{5/4}
    \int_{T_{min}}^{T_{kin}} \vev{\sigma v}
    \calA_1 g_1 ~\calA_2 g_2 \,M dT 
\nonumber \\ &\equiv
    Y_{\Gt}^{(1)}(\tthr) + Y_{\Gt}^{(2)}(\tthr) \,,
\label{eq:Gtthrabund}
\end{align}
where $Y_\Gt (\tthr) = n_\Gt(\tthr)/s(\tthr)$ and $s(\tthr) = (2\pi^2/45)
g_* T_R^3$.  $Y_{\Gt}^{(1)}$ and $Y_{\Gt}^{(2)}$ are associated with
gravitino production between $t_d$ and $\tkin$, and between $\tkin$ and 
$\tthr$, respectively.

For $Y_{\Gt}^{(2)}(\tthr)$,
substituting \eqs{\eqref{eq:sigmav}} in \eq{\eqref{eq:Gtthrabund}} we have
\begin{align}
Y_{\Gt}^{(2)}(\tthr) &= 
    \left[
    \frac{0.07}{g_*^{3/2}}   
    \left(\frac{g_*}{\calA} \right)^{5/4}
    \sum_l f_l \,\frac{\alpha}{32 M^2} \,\left(1+\frac{m_0^2}
    {3 m_\Gt^2} \right)
    \,\calA_1 g_1 \, \calA_2 g_2 M \, \right]
   \int_{T_{min}}^{T_{kin}} dT 
\nonumber \\ &=
    \left[
    \frac{0.07}{g_*^{3/2}}   
    \left(\frac{g_*}{\calA} \right)^{5/4}
    \sum_l f_l \,\frac{\alpha}{32 M^2} \,\left(1+\frac{m_0^2}
    {3 m_\Gt^2} \right)
    \,\calA_1 g_1 \, \calA_2 g_2 M \, \right]
    \left(T_{kin} - T_{min} \right)
\end{align}
where $f_l = 48, 16, 8$ from \eq{\eqref{eq:sigmav}} for
$l = E, I, J$ processes respectively and $g_{1,2}$ are as given in
\ref{sec:bparam}. We have ignored the variation of $\alpha$ with
temperature and shall use the value of $\alpha = 5 \times 10^{-2}$.

\subsubsection{$t > \tthr$}

As discussed in \sect{\ref{ssec:tm0gGd}} gravitino production is
suppressed until the flat direction condensate decays at $t_f$. The
temperature $T_f$ at $t_f$ is given in \sect{\ref{sec:Tf}} and equals
$1 \times 10^5 \GeV$. Such a low temperature will lead to a gravitino
abundance consistent with cosmological constraints.

\section{Obtaining parameters}
\label{sec:params}

The values of parameters which we require to obtain the gravitino
abundances are evaluated in this section for both $m_0 < \Gamma_d$
and $m_0 > \Gamma_d$.

\subsection{$T_{kin}$ and $\Gamma_{kin}$}

$T_{kin}$ is the temperature at $\tkin$ just after kinetic equilibration.
Since kinetic equilibration only re-distributes energy among particles
without changing their number, the average energy per particle is the same
before and after kinetic equilibration. Therefore
\begin{align}
\frac{3}{2} \,k_B \,T_{kin} 
&=
    \frac{m_\phi}{2}\frac{a_d}{a_{kin}} 
\nonumber \\ &=
    \frac{m_\phi}{2}\left(\frac{t_d}{t_{kin}}\right)^{1/2} 
\nonumber \\ &=
    \frac{m_\phi}{2}\left(\frac{\Gamma_{kin}}{\Gamma_d}\right)^{1/2} ,
\end{align}
where the RHS represents the energy of the particles just before kinetic
equilibration and we presume 2 body decay of the inflaton. The Boltzmann
constant $k_B$ is hereafter set to $1$.  Therefore
\begin{equation}
T_{kin} = \frac{m_\phi}{3} 
\left( \frac{\Gamma_{kin}}{\Gamma_{d}}
\right)^{1/2}\, , 
\label{eq:Tkin}
\end{equation}
where $\Gamma_{kin}$  is given in \eqs{\eqref{eq:gthr-m0lGd}} and
\eqref{eq:Gtthrm0gGd}.

For $m_0 < \Gamma_d$, 
\begin{align}
\Gamma_{kin} &=
    \frac{3}{2} \frac{\alpha \,M_{Pl}^2 \,\Gamma_d^{1/2} \,m_0^{3/2}}
    {m_\phi \,\varphi_0^2}
    = 11 \GeV\,.
\label{eq:Gkinm0lGd}
\end{align}
For $m_0 > \Gamma_d$, 
\begin{align}
\Gamma_{kin} &= 
    \frac{3}{2} \frac{\alpha \,M_{Pl}^2 \,m_0^2}{m_\phi \,\varphi_0^2}.
\nonumber \\ &
    = 1 \GeV\,.
\label{eq:Gkinm0gGd}
\end{align}
Thus, the value of $T_{kin}$ for $m_0 < \Gamma_d$ is $1 \times
10^{11}\GeV$ and for $m_0 > \Gamma_d$ is $1 \times 10^{12} \GeV$.

\subsection{$T_{min}$}
\label{ssec:tmin}

The temperature $T_{min}$ just before thermalization at $\tthr$ can be
obtained by
\begin{align}
T_{min} &= T_{kin} \left( \frac{a_{kin}}{a_{thr}} \right) 
    = T_{kin} \left( \frac{t_{kin}}{t_{thr}} \right)^\half
    = T_{kin} \left( \frac{\Gamma_{thr}}{\Gamma_{kin}} \right)^\half
\label{eq:Tminkin}
\\ &= 
    T_{kin} \sqrt{\alpha} 
\nonumber \\ T_{min} &=
    \begin{cases}
        2 \times 10^{10} \GeV & \textrm{for } m_0 < \Gamma_d \\
        2 \times 10^{11} \GeV & \textrm{for } m_0 > \Gamma_d 
    \end{cases}
\end{align}
where we have used the relation $\Gamma_{thr} = \alpha \Gamma_{kin}$ as
discussed in \sect{\ref{sec:discuss}}.

\subsection{$\calA$} 
\label{ssec:calA}

To obtain $\calA$ we note that the energy density of the Universe 
$\rho_{kin^+}$ just after kinetic equilibration at $\tkin$ is equal to 
the energy density $\rho_{kin^-}$ just before kinetic equilibration. Then
\begin{align}
\rho_{kin^+} &= 
    \rho_{kin^-}
\nonumber \\
    \textrm{or, } \qquad \frac{\pi^2}{30} \calA T_{kin}^4  
&= 
    \rho_\phi(t_d) 
    \left( \frac{a_d}{a_{kin}} \right)^4
\nonumber \\ &=
    \rho_{\phi}(t_d) \left( \frac{t_d}{t_{kin}} \right)^2 
\nonumber \\
    \textrm{or, } \qquad 
    \frac{\pi^2}{30}\calA
    \left( \frac{m_{\phi}}{3} 
    \left( \frac{\Gamma_{kin}}{\Gamma_d} \right)^\half  \right)^4 
&=
    \rho_{\phi}(t_d) \left( \frac{\Gamma_{kin}}{\Gamma_d}
    \right)^2 
\nonumber \\
    \textrm{or, } \qquad \calA &\approx
    \frac{2430}{\pi^2} \left(\frac{\rho_{\phi}(t_d)}{m_\phi^4} \right)
\nonumber \\
    &\approx \frac{2430}{\pi^2} \left( \frac{3}{8\pi}\right) 
    \left(\frac{M_{Pl}^2 \Gamma_d^2 }{m_\phi^4} \right)
\nonumber \\
    \textrm{or, } \qquad
    \calA &\sim 
        \begin{cases}
        4 \times 10^{-5}, & \textrm{ for }
            m_0 < \Gamma_d = 10^4 \textrm{ GeV} \\
        4 \times 10^{-11}, & \textrm{ for } 
            m_0 > \Gamma_d = 10 \textrm{ GeV}.
        \end{cases}
\end{align}
Thus $\calA \ll 1$. 

\subsection{$A_i$}
\label{ssec:Ai}

The number density of a species $i$ just after kinetic equilibration is
$\calA_i g_i T_{kin}^3/\pi^2$. This equals the number density of the
species just before kinetic equilibration which, in turn, is the number 
density of the species at $t_d$ scaled by $1/a^3$. The decay rate of
the inflaton to a species $i$ will be proportional to $g_i$ and so 
$n_i (t_d)$ is also proportional to $g_i$. Then $\calA_i$ is the same
for all species. Since $\calA = \sum_j \calA_j g_j = \calA_i \sum_j 
g_j \approx 200 \, \calA_i $, $\calA_i \approx \calA/200$.

\subsection{$T_R$}
\label{ssec:TR}

As stated earlier in \eq{\eqref{eq:TrTmin}} the temperature at $\tthr$ is
given by
\begin{align}
T_R &= \left( \frac{\calA}{g_*} \right)^{1/4} T_{min}
\nonumber \\ &=
    \left( \frac{\calA}{g_*} \right)^{1/4} \sqrt{\alpha} \,T_{kin}
\nonumber \\ &= 
    \left( \frac{\calA}{g_*} \right)^{1/4} \sqrt{\alpha} 
    \,\frac{m_\phi}{3} 
    \paren{\frac{\Gamma_{kin}}{\Gamma_d}}^{1/2}.
\label{eq:TRcommon}
\end{align}
For $m_0 < \Gamma_d$,
\begin{align}
T_R &=
    \frac{1}{\sqrt{6}}
    \left( \frac{\calA}{g_*} \right)^{1/4}  
    \frac{\alpha \,m_\phi^{1/2} \,M_{Pl} \,m_0^{3/4}}
    { \,\varphi_0 \,\Gamma_d^{1/4}}
\nonumber \\ &=
    5 \times 10^8 \GeV \,.
\label{eq:TRm0lGd}
\end{align}
For $m_0 > \Gamma_d$,
\begin{align}
T_R &=
    \frac{1}{\sqrt{6}}
    \left( \frac{\calA}{g_*} \right)^{1/4}  
    \frac{\alpha \,m_\phi^{1/2} \,M_{Pl} \,m_0}
    { \,\varphi_0 \,\Gamma_d^{1/2}}
\nonumber \\ &=
    2 \times 10^8 \GeV \,.
\label{eq:TRm0gGd}
\end{align}
We have used \eqs{\eqref{eq:Gkinm0lGd}} and \eqref{eq:Gkinm0gGd}
for the expression of $\Gamma_{kin}$ for $m_0 < \Gamma_d$ and 
$m_0 > \Gamma_d$ respectively, and $\alpha = 5 \times 10^{-2}$.

\subsection{$T_G$}
\label{ssec:TG}

The temperature $T_G$ at $t_g$ is given by
\begin{align}
T_G &= T_R \paren{\frac{a_{thr}}{a_G}}\,.
\end{align}
For $m_0 < \Gamma_d$, using \eq{\eqref{eq:aGm0lGd}} to define $a_G$, we get
\begin{align}
T_G &=
    \frac{T_R^3}{\alpha \,\varphi_0^2} \paren{m_0 \,t_{thr}}^{3/2}
\nonumber \\ &=
\paren{\frac{1}{27}} \paren{ \frac{\calA}{g_*} }^{3/4} 
	\frac{m_\phi^3 \,m_0^{3/2}}{\alpha \,\varphi_0^2
    \,\Gamma_d^{3/2}} 
\nonumber \\ &=
    300 \GeV \,,
\label{eq:TGm0lGd}
\end{align}
where we have used \sect {\ref{sssec:eqtimes}} for $\tthr$ and 
\eq \eqref{eq:TRcommon} or \eqref{eq:TRm0lGd} for $T_R$, and $\alpha = 0.1$.

For $m_0 > \Gamma_d$, using \eq{\eqref{eq:aGm0gGd}}, the temperature at 
$t_G$ is 
\begin{align}
T_G &=     
    \frac{T_R^3}{\alpha \,\varphi_0^2} \paren{\frac{m_0}{\Gamma_d}}^2
    \paren{\Gamma_d \,\tthr}^{3/2}
\nonumber \\ &=
    \frac{1}{27} \paren{ \frac{\calA}{g_*} }^{3/4} 
	\frac{m_\phi^3 \,m_0^2}{\alpha \,\varphi_0^2
    \,\Gamma_d^2} 
\nonumber \\ &=
    1000 \GeV \,,
\label{eq:TGm0gGd}
\end{align}
where we have used \sect \ref{sssec:noeqcond-m0gGd} for $\tthr$ and \eq
\eqref{eq:TRcommon} or \eqref{eq:TRm0gGd} for $T_R$, and $\alpha = 0.1$.

\subsection{$T_f$}
\label{sec:Tf}

The temperature at $t_f$ is given by 
\begin{align}
T_f &= T_R (\tthr/t_f)^{1/2}\,.
\end{align}
For $m_0 < \Gamma_d$,
\begin{align}
T_f &=  \frac{1}{3} \paren{\frac{\calA}{g_*}}^{1/4} 
    \frac{m_\phi \,m_0^{9/10}}{\varphi_0^{2/5} \, \Gamma_d^{1/2}}
\nonumber \\ &=
    1 \times 10^5 \GeV \,,
\label{eq:Tfm0lGd}
\end{align}
where the expressions for $\tthr$, $t_f$ and $T_R$ are obtained from
\sect \ref{sssec:eqtimes}, \eqs \eqref{eq:tfm0lGd} and \eqref{eq:TRcommon}
or \eqref{eq:TRm0lGd} respectively.

For $m_0 > \Gamma_d$,
\begin{align}
T_f &= \frac{1}{3} \paren{\frac{\calA}{g_*}}^{1/4} 
    \frac{m_\phi \,m_0}{\varphi_0^{2/5} \, \Gamma_d^{3/5}}
\nonumber \\ &=
    1 \times 10^5 \GeV \,,
\label{eq:Tfm0gGd}
\end{align}
where the expressions for $\tthr$, $t_f$ and $T_R$ are obtained from 
\sect \ref{sssec:noeqcond-m0gGd}, \eqs \eqref{eq:tfm0gGd} and 
\eqref{eq:TRcommon} or
\eqref{eq:TRm0gGd} respectively.

\section{Results}
\label{sec:results}

For the case $m_0 < \Gamma_d$, with $m_0 = 100 \GeV$, $\Gamma_d =
10^4 \GeV$ and $\varphi_0 = 10^{14} \GeV$, $t_d$, $t_0, \tkin$,
$\tthr$ and $t_f$ are $10^{-4} \GeV^{-1}$, $10^{-2}\GeV^{-1}$, $9
\times 10^{-2}\GeV^{-1}$, $2 \GeV^{-1}$ and $4 \times 10^7 \GeV^{-1}$
respectively. $\varphi$ is constant at $10^{14} \GeV$ from $t_d$
to $t_0$ and falls to $2 \times 10^{13} \GeV$ and $2 \times 10^{12}
\GeV$ at $\tkin$ and $\tthr$ respectively. From \ref{sec:mgt}, the
gluon mass is constant from $t_d$ to $t_0$ at $2 \times 10^{13} \GeV$
and falls to $4 \times 10^{12}$ and $4 \times 10^{11} \GeV$ at $\tkin$
and $\tthr$ respectively.  The final reheat temperature $T_R$ after full
equilibration is $5\times10^8\GeV$, which is less than the gluon and
gluino mass at that time. As we have argued in \sect{\ref{ssec:tm0lGd}}
there is no gravitino production in the non-thermal phase, nor after
thermalization till the flat direction condensate decays at $t_f$,
because of phase space suppression associated with the large mass of the
outgoing gluon or gluino. The temperature $T_f$ is $1 \times 10^5 \GeV$
and gravitino production after $t_f$ is hence small and consistent with
cosmological constraints.

For $m_0 > \Gamma_d$, with $m_0 = 100 \GeV$, $\Gamma_d = 10 \GeV$ and
$\varphi_0 = 10^{14} \GeV$, $t_0$, $t_d$, $\tkin$ and $\tthr$ are $0.01
\GeV^{-1}$, $0.1 \GeV^{-1}$, $0.9 \GeV^{-1}$ and $20 \GeV^{-1}$. $\varphi$
falls to $10^{13} \GeV$ at $t_d$, $2 \times 10^{12} \GeV$ at $t_{kin}$
and $2 \times 10^{11} \GeV$ at $t_{thr}$. The corresponding gluon mass
is $2 \times 10^{13} \GeV$, 
$2 \times 10^{12} \GeV$, $4 \times 10^{11}\GeV$ and
$4 \times 10^{10}\GeV$ at $t_0$, $t_d$, $\tkin$ and $\tthr$ respectively. The
final reheat temperature $T_R$ is $2\times10^8 \GeV$.  The gravitino
abundances generated in the different epochs are
\begin{align}
Y_{\Gt}^{(1)} &= 
4 \times 10^{-18}
\nonumber \\
Y_{\Gt}^{(2)} &= 
1 \times 10^{-20}
\end{align}
where $Y_{\Gt}^{(1)}$ is the abundance due to gravitino production
between $t_d$ and $\tkin$ and $Y_{\Gt}^{(2)}$ is the abundance due to
production between $\tkin$ and $\tthr$.  The gravitino abundance in
both epochs is less than the upper bound of $10^{-14}$ on the gravitino
abundance from various cosmological constraints \cite{Cyburt:2009pg}.
Gravitino production after thermalization at $\tthr$ is suppressed
until the condensate decays at $t_f$. Since the temperature at $t_f$
is $1 \times 10^5 \GeV$ the gravitino abundance generated after $t_f$
will be small and within cosmological bounds.

\section{Conclusions}
\label{sec:conclusions}

In this article we have studied the novel scenario proposed by \refes
\cite{Allahverdi:2005fq,Allahverdi:2005mz} where the vacuum expectation
values of SUSY flat directions generate a large mass for gauge bosons and
gauginos. Due to the large gauge boson masses thermalization is delayed
after the end of inflation.  In the non-thermal Universe gravitino
production in \refes \cite{Allahverdi:2005fq, Allahverdi:2005mz} is
suppressed because of the dilute nature of the plasma. The delay in
thermalization also leads to a low reheat temperature and so gravitino
production after thermalization is also suppressed. Thus it is argued
that the gravitino problem of SUSY models is avoided in this scenario.

In this article we have carried out a more careful analysis of
gravitino production in the non-thermal Universe scenario of \refes
\cite{Allahverdi:2005fq,Allahverdi:2005mz}.  We have identified the
initial state particle distribution functions in the gravitino production
process for the eras before kinetic equilibration and after kinetic
equilibration in a non-thermal Universe (the appropriate particle
distribution function for the pre-kinetic equilibration era was not
used in \refes  \cite{Allahverdi:2005fq,Allahverdi:2005mz}).  These have
then been included in the calculation of the collision integral in the
integrated Boltzmann equation.  We have also explicitly obtained the
conditions for obtaining a non-thermal Universe after inflaton decay.
We have further investigated final state phase space suppression in the
context of heavy final state gluons and gluinos in both the non-thermal
and thermal eras.

We have investigated two cases (i) where the flat direction starts
oscillating after inflaton decay ($m_0 < \Gamma_d$) and (ii) where
the flat direction starts oscillating before inflaton decay ($m_0 >
\Gamma_d$).  In the first case ($m_0 < \Gamma_d$) we find that there is no
gravitino production during the non-thermal phase because of phase space
suppression due to the large gluon and gluino mass. Gravitino production
is suppressed even after thermalization till the flat direction condensate
decays at $t_f$.  However the temperature $T_f$ is low and as such any
gravitino abundance produced after $t_f$ is small and is within the
standard cosmological bounds.

In the second case ($m_0 > \Gamma_d$) there is gravitino production before
thermalization.  The gravitino abundance generated is consistent with
cosmological constraints.  After thermalization, as in the case $m_0 <
\Gamma_d$, there is no gravitino production before the flat direction
condensate decays.  After the decay of the condensate the temperature
$T_f$ is low and does not lead to conflict with cosmological bounds.

\section*{Acknowledgments} 
We would like to thank H. Baer, A. Joshipura, P. Konar, N. Mahajan, H. Mishra,
K. Patel and P. Roy for very useful discussions. We would also like to thank
R. Allahverdi for clarifying aspects of \refe \cite{Allahverdi:2005mz}.

\appendix

\section{Normalization of the phase space distribution functions for 
$t_d < t < \tkin$}
\label{sec:phspace} 

The number of incoming particles $N_i$ for gravitino production is given by
\begin{equation}
\frac{g_i}{(2\pi)^3} \int f_i ~d^3p_i ~dV = N_i .
\end{equation}
For the time interval $\td < t < \tkin$ the form of $f$ is given by 
\eq{\eqref{eq:fi1}}. Substituting that in the above equation we get,
\begin{align}
N_i/V \equiv n_i(t) &=
    \frac{g_i}{(2\pi)^3} \int C_i 
    \delta \left(E_i - \frac{m_\phi}{2} \frac{a_d}{a} \right)
    4 \pi p_i^2 ~dp_i 
\nonumber \\ &=
    \frac{g_i}{(2\pi)^3} \int C_i 
    \delta\left(E_i - \frac{m_\phi}{2} \frac{a_d}{a} \right)
    4 \pi \sqrt{E_i^2 - m_i^2} ~E_i ~dE_i 
\nonumber \\ \textrm{or, } \qquad
C_i &= 
    \frac{n_i(t)}{m_\phi^2} \left( \frac{a}{a_d}\right)^2   
    \frac{(2\pi)^3}{\pi g_i}.
\label{eq:cint} 
\end{align}
where we have considered $E_i \gg m_i$. Now at $t = \td$, the
number density of the particles is equal to the number density of the
inflaton, as $\phi \rightarrow X_1 + X_2$. So, $n_i(\td) = n_\phi(\td) =
\rho_\phi(t_d)/m_\phi$. Then for $\td < t < \tkin$, the number density
$n_i(t)$ is given by $n_i(t) = (\rho_\phi(t_d)/m_\phi) ( a_d/a)^3 $.
Substituting the above expression for $n_i(t)$ in \eq{\eqref{eq:cint}} we get
\begin{equation}
C_i(t) = \frac{\rho_\phi (\td)}{m_\phi^3} \left( \frac{a_d}{a}\right)   
    \frac{(2\pi)^3}{\pi g_i}.  
\label{eq:cideriv} 
\end{equation}

\section{$A$ terms for $\td < t < \tkin$}
\label{sec:Aterms}

Below we present the $A$ terms for the processes E, I and J during the
interval $\td < t < \tkin$ as needed for the $m_0 > \Gamma_d$ case using
\eqs{\eqref{eq:mbltzeq}}, \eqref{eq:w12cm}, \eqref{eq:p12} and 
\eqref{eq:dP1dP2}, and the
phase space distribution in \eq{\eqref{eq:fi1}}, with the expressions
for $\sigma_{CM}$ for the $\pm 1/2$ helicity gravitino states as given in
\eqs{\eqref{eq:sigmae}}, \eqref{eq:sigmai} and \eqref{eq:sigmaj}.

The A term for the E process is
\begin{align}
A_E(t) &= 
\frac{ 
6 \times 2 \times 2 \times 
\sum_{A,i,j} \,C_1 \,C_2 \,g_s^2\,|T^A_{ji}|^2   
\,a_d^4 \,m_{\phi}^{4} }{1024  \,M^2 \,\pi ^5 \,a^4 }
\left( 1 + \frac{m_0^2}{3 m_\Gt^2}\right)
\,,
\end{align}
where $T^A$ is the generator of the $SU(3)$ group in the
fundamental representation and $C_i \propto (a_d/a)$ are given in
\eq{\eqref{eq:cideriv}}.  $A_E$ includes a factor of $6$ for $6$ quark/squark
flavors, a factor of $2$ for the $2$ squark `chiralities' and another 
factor of $2$ for the $2$ hermitian conjugate processes. 
(One should actually exclude the heavy quarks and squarks associated with
the flat direction.  That would decrease the final abundance by a factor
of $O(1)$.) Then we get
\begin{equation}
A_E(t) =
144 B 
m_{\phi}^4 \left(\frac{a_d^6}{a^6}\right) ,
\end{equation}
where the parameter $B$ is as defined in \ref{sec:bparam}. 

The A term for the I process is
\begin{align}
A_I(t) &= \frac{ 
6 \times
\sum_{A,i,j} \,C_1 \,C_2 \,g_s^2 \,|T^A_{ji}|^2 
\,a_d^4 \,m_{\phi }^4} {768 \,M^2 \,\pi ^5 \,a^4}
\left( 1 + \frac{m_0^2}{3 m_\Gt^2}\right)
\,. 
\end{align}
$A_I$ includes a factor of $6$ for the quark flavors.  Then we get
\begin{equation}
A_I(t) = 48 B 
\, m_{\phi }^4 \left( \frac{a_d^6}{a^6} \right).
\end{equation}

The A term for the J process is
\begin{align}
A_J(t) &= 
\frac{ 
6 \times 2 \times
\sum_{A,i,j} \,C_1 \,C_2 \,g_s^2 \, |T^A_{ji}|^2 
    \,a_d^4 m_{\phi }^{4}}
    {768 \, M^2 \,\pi ^5  \,a^4} 
\left( 1 + \frac{m_0^2}{3 m_\Gt^2}\right)
\,.
\end{align}
$A_J$ gets a factor of $6$ for squark flavors and a factor of $2$ for the
squark `chiralities'.  Then we get
\begin{equation}
A_J(t) = 96 B 
m_{\phi }^4 \left(\frac{a_d^6}{a^6} \right) .
\end{equation}

\section{B parameter}
\label{sec:bparam} 

To simplify our equations we define new parameters called $\cal{B}$ and 
$B$ for the pre-factor present in $A$ terms, i.e.,
\begin{equation}
\mathcal{B} = \left( \frac{ \sum_{A,i,j} 8 \pi \,C_1 \,C_2 \,g_s^2
\,|T^A_{ji}|^2 } {6144 \,M_{Pl}^2 \,\pi^5} \right) 
\left( 1 + \frac{m_0^2}{3 m_\Gt^2}\right) 
\,,
\end{equation}
where $M_{Pl}^2 = 8 \pi M^2$. Substituting the values of $C_1$ and $C_2$
from \eq{\eqref{eq:cideriv}} in the above equation gives us
\begin{align}
\mathcal{B} &= \left[ \left( \frac{\rho_\phi (\td)}{m_\phi^3} 
    \frac{(2\pi)^3}{\pi} \right)^2 \left(\frac{1}{g_1 g_2} \right) 
    \right] \paren{
    \frac{ \sum_{A,i,j} 8\pi\,g_s^2 \,|T^A_{ji}|^2}
    {6144 \,M_{Pl}^2 \,\pi^5 } }
    \left( 1 + \frac{m_0^2}{3 m_\Gt^2}\right) 
    \left( \frac{a_d}{a} \right)^2 
\equiv B \left( \frac{a_d}{a} \right)^2 ,
\end{align}
where 
\begin{align}
B &= \left[\left( \frac{\rho_\phi (\td)}{m_\phi^3} 
    \frac{(2\pi)^3}{\pi} \right)^2 \paren{\frac{1}{g_1 g_2}} 
    \right] \paren{
    \frac{ \sum_{A,i,j} 8\pi \,g_s^2 \,|T^A_{ji}|^2}
    {6144 \,M_{Pl}^2 \,\pi^5} }
    \left( 1 + \frac{m_0^2}{3 m_\Gt^2}\right) 
\nonumber \\ &=
    \left[\left( \frac{3 \,\Gamma_d^2}{m_\phi^3} 
    \frac{(2\pi)^3}{\pi} \right)^2 \paren{\frac{M_{Pl}^2}{8\pi}} 
    \right] \paren{
    \frac{ \sum_{A,i,j} \,g_s^2 \,|T^A_{ji}|^2 }
    {6144 \,g_1\,g_2 \,\pi^5 } }
    \left( 1 + \frac{m_0^2}{3 m_\Gt^2}\right) 
\nonumber \\ &=
    \frac{ 3 \sum_{A,i,j} g_s^2 \,M_{Pl}^2 \,|T^A_{ji}|^2  \,\Gamma_d^4}
    {256 \,\pi^2 \,g_1\,g_2 \,m_\phi^6} 
    \left( 1 + \frac{m_0^2}{3 m_\Gt^2}\right) 
\label{eq:bparam}
\end{align}
and we have used $\rho_\phi(t_d) = (3/ (8\pi)) M_{Pl}^2 \,\Gamma_d^2$.
We suppress the index $E,I,J$ for $B$ associated with the different
$g_{1,2}$ for each process.  Considering the spin and color of the
incoming particles, $(g_1,g_2) = (6,3),(6,6)$ and $(3,3)$ for the $E,I,J$
processes respectively.

\section{Mass of the gluon}
\label{sec:mgt}

The mass of the gluon  $m_{g} = \sqrt{\alpha}\varphi$ which 
varies
as $1/a^{3/2}$ after the condensate starts oscillating at $t_0$. Below
we derive the mass of the gluon 
for various epochs in time
for the two cases, $m_0 < \Gamma_d$ and $m_0 > \Gamma_d$.

\subsection{$m_0 < \Gamma_d$ or $t_0 > \td$}

\begin{enumerate}

\item For $t_d < t < t_0$, $\varphi = \varphi_0$. So
\begin{equation}
m_{g}(t_d) = \sqrt{\alpha} \varphi_0 \,.
\label{eq:Imgttd}
\end{equation}

\item For $t_0 < t < \tthr$,
\begin{align}
m_{g} &= m_{g}(t_0) \paren{\frac{a_0}{a}}^{3/2} 
\nonumber \\ &=
    \sqrt{\alpha} \varphi_0 \paren{\frac{t_0}{t}}^{3/4} 
\nonumber \\ &= 
    \sqrt{\alpha} \varphi_0 \paren{\frac{1}{m_0 \,t}}^{3/4} \,.
\end{align}
At $t = \tkin$,
\begin{align}
m_{g} (\tkin) &=
    \sqrt{\alpha} \varphi_0 \paren{\frac{\Gamma_{kin}}{m_0}}^{3/4} \,,
\end{align}
where $\Gamma_{kin}$ is given by \eq{\eqref{eq:Gkinm0lGd}}.

At $t = \tthr$,
\begin{align}
m_{g} (\tthr) &=
    \sqrt{\alpha} \varphi_0 \paren{\frac{\Gamma_{thr}}{m_0}}^{3/4} \,. 
\label{eq:Imgttthr}
\end{align}

\item For $t > \tthr$,
\begin{align}
m_{g} &= m_{g}(\tthr) \paren{ \frac{a_{thr}}{a} }^{3/2}
\nonumber \\ &=
    \sqrt{\alpha} \varphi_0 \paren{\frac{\Gamma_{thr}}{m_0}}^{3/4}
\paren{\frac{T}{T_{R}}}^{3/2} \,.
\label{eq:Imgttf}
\end{align}
\end{enumerate}

\subsection{$m_0 > \Gamma_d$ or $t_0 < \td$ }

\begin{enumerate}

\item At $t = t_0$,
\begin{equation}
m_{g}(t_0) = \sqrt{\alpha} \varphi_0 \,.
\end{equation}

\item For $t_0 < t < t_d$,
\begin{align}
m_{g} &= 
    \sqrt{\alpha} \varphi_0 \paren{\frac{a_0}{a}}^{3/2}
\nonumber \\ & =
    \sqrt{\alpha} \varphi_0 \paren{\frac{t_0}{t}}   
\nonumber \\ &= 
\sqrt{\alpha} \varphi_0 \paren{\frac{1}{m_0 \,t}},
\end{align}
where we have used the fact that the Universe was matter dominated for the
period $t_0 < t < t_d$ and $a \sim t^{2/3}$, as relevant for an inflaton
oscillating in a quadratic potential after inflation.

At $t = t_d$,
\begin{equation}
m_{g}(t_d)=
    \sqrt{\alpha} \varphi_0 \paren{\frac{\Gamma_d}{m_0}},
\label{eq:IImgttd}
\end{equation}

\item For $t_d < t < t_{kin}$,
\begin{align}
m_{g} &= 
    m_{g}(t_d) \paren{\frac{a_d}{a}}^{3/2}
\nonumber \\ &=
    m_{g}(t_d)
    \paren{\frac{t_d}{t}}^{3/4}
\nonumber \\ &= 
    \sqrt{\alpha} \varphi_0 \paren{\frac{\Gamma_d}{m_0}}
    \paren{\frac{1}{\Gamma_d \,t}}^{3/4},
\label{eq:IImgttkin}
\end{align}
where we have used the fact that the Universe is radiation dominated for the
period $t_d < t < t_{kin}$ and $a \sim t^{1/2}$.

\item For $t_{kin} < t < t_{thr}$
\begin{align}
m_{g} &= 
    m_{g}(t_{kin}) \paren{\frac{a_{kin}}{a}}^{3/2}
\nonumber \\ &=
    \sqrt{\alpha} \varphi_0 
    \paren{\frac{\Gamma_d}{m_0}}
    \paren{\frac{\Gamma_{kin}}{\Gamma_d}}^{3/4}
    \paren{\frac{T}{T_{kin}}}^{3/2} \,,
\label{eq:IImgttthr}
\end{align}
where we have used the relation $a \sim 1/T$ for $t_{kin} < t < t_{thr}$.

\item For $t > t_{thr}$
\begin{align}
m_{g}(t) &= m_{g}(t_{thr}) \paren{\frac{a_{thr}}{a}}^{3/2}
\nonumber \\ &=
    \sqrt{\alpha} \varphi_0 
    \paren{\frac{\Gamma_d}{m_0}}
    \paren{\frac{\Gamma_{thr}}{\Gamma_d}}^{3/4}
    \paren{\frac{T}{T_R}}^{3/2} \,,
\label{eq:IImgttf}
\end{align}
where we have used \eqs{\eqref{eq:IImgttthr}} and \eqref{eq:Tminkin},
and $a \sim 1/T$ for $t > \tthr$. 
\end{enumerate}

% \section*{References}
% %\bibliographystyle{elsarticle-num}
% \bibliographystyle{JHEP}
% \bibliography{cosmology,susy,genesis,books,qballs}
% \end{document}

\providecommand{\href}[2]{#2}\begingroup\raggedright\endgroup

\end{document}